\documentclass[twocolumn,times,twocolappendix]{aastex631}

\usepackage{amsmath}
\usepackage[T1]{fontenc}

\newcommand\ergs{erg~s$^{-1}$}

\newcommand{\halpha}{H{$\alpha$}}

\newcommand{\HeII}{He\,{\sc ii}}

\newcommand{\SII}{[S\,{\sc ii}]}

\shorttitle{Optical emission from luminous and very soft X-ray sources}
\shortauthors{Tang \& Feng}

\graphicspath{{./}{fig/}}

\begin{document}

\title{Optical emission from luminous and very soft X-ray sources in nearby galaxies: testing the scenario of edge-on supercritical accretion systems}

\author{Xiaohong Tang}
\affiliation{Department of Astronomy, Tsinghua University, Beijing 100084, China}

\author[0000-0001-7584-6236]{Hua Feng}
\email{hfeng@ihep.ac.cn}
\affiliation{Key Laboratory of Particle Astrophysics, Institute of High Energy Physics, Chinese Academy of Sciences, Beijing 100049, China} 

\begin{abstract}

Supercritical accretion onto compact objects is expected to drive optically thick winds, resulting in observed X-ray emission as a function of viewing angle. 
However, their optical emission, either from the outer accretion disk or companion surface tends to be nearly isotropic.
Based on a sample of luminous and very soft X-ray sources that are argued to be supercritical accretion systems viewed close to edge-on, we identified the optical counterparts for some of them and compared the optical properties with those of ultraluminous X-ray sources (ULXs), which are supposed to be supercritical accretion systems viewed close to face-on.  
The optical luminosity is found in a wide range, with the absolute visual magnitude ranging from dimmer than $-1.2$ in some sources to about $-7$ in one case. 
Most sources show a power-law like spectrum while four of them display a blackbody shape. 
One of them shows an optical spectrum resembling a B type main sequence, suggesting that it may be a Be white dwarf system. 
Strong variability in flux at timescales as short as 10 days are revealed, indicating that some of these sources are powered by accretion onto compact objects. 
These suggest that the luminous and very soft X-ray sources in nearby galaxies have a diverse population, and some of them are indeed consistent with emission from supercritical accretion, with consistent optical magnitudes and colors. 
Future optical spectroscopic observations are needed to further constrain their natures. 
\end{abstract}

\section{Introduction}
\label{sec:intro}

Supercritical accretion onto stellar compact objects may drive ultraluminous X-ray sources (ULXs) with an X-ray luminosity higher than $10^{39}$~\ergs\ \citep{ULXreview}. 
However, the observed X-ray flux could be low due to obscuration by the massive disk wind that is inevitable in extreme accretion \citep{King2023,Pinto2023}. 
A remarkable example is SS~433, which harbors a stellar mass compact object accreting from an evolved massive donor star with a mass accretion rate $\sim$$10^2$ times the Eddington level, but the observed X-ray emission from the central core is rather faint \citep{Fabrika2007}. 

The massive wind is expected to be optically thick and manifests itself as a low-temperature thermal component dominant in the soft X-ray or UV band \citep{Meier1982III, Poutanen2007}. 
Numerical simulations revealed that the massive wind encloses a central low-density funnel, where the hard X-ray emission from the central accretion flow may escape \citep{2014ApJ...796..106J,2016MNRAS.456.3929S,2017MNRAS.469.2997N}.
In this picture, systems viewed close to the symmetric axis appear as standard ULXs with a significant hard X-ray component \citep{Middleton2015}, while those viewed at large inclination angles tend to be soft \citep{Pinto2016}.
Observationally, the supersoft ULXs show behaviors in agreement with the wind scenario \citep{Feng2016,Soria2016,2016MNRAS.456.1859U}, and the soft excess seen in the spectrum of standard ULXs is also consistent with such an origin \citep{Qiu20212}.

As the wind is driven by radiation pressure, its luminosity on the photosphere is in principle capped at the Eddington limit \citep{Meier1982III}. 
ULXs contain both stellar mass black holes and neutron stars \citep{Bachetti2014}. 
Therefore, the wind luminosity is not necessarily ultraluminous and could be as low as $\sim$$10^{38}$~\ergs\ in the case of a neutron star accretor \citep{Qiu20212}. 
This suggests that one may find edge-on supercritically accreting compact objects in non-ULXs. 
Triggered by this idea, \citet{Zhou2019} conducted a search of such candidates in nearby galaxies in the archive of the Chandra X-ray Observatory. 
These sources show a spectrum dominated by a luminous ($L_{\rm bb} \approx 10^{37} - 10^{40}$~\ergs)\footnote{A luminosity down to $10^{37}$~\ergs\ is for the consideration of observational uncertainties.} and very soft ($kT \approx 0.05 - 0.4$~keV) blackbody component, and are ruled out to be foreground stars or supernova remnants based on optical emission.
If they are indeed supercritical accretion systems viewed edge-on, their optical emission should show properties comparable to those of ULXs. 

Thus in this study, we identified the optical counterpart of sources in the \citet{Zhou2019} catalog using images in the Hubble Space Telescope (HST) archive (\S~\ref{sec:sample}), compared them with the standard ULXs in terms of optical properties, and discussed their possible physical natures (\S~\ref{sec:discuss}).

\tabletypesize{\scriptsize}
\begin{deluxetable*}{ccccccccc}
\tablecaption{Sample of luminous and very soft X-ray sources studied in this work.}
\label{tab:sample}
\colnumbers 
\tablehead{
\colhead{ID} & \colhead{Host}   & \colhead{corrected R.\ A.} & \colhead{corrected Decl.}&\colhead{$\delta$}& \colhead{$d_{\rm L}$} & \colhead{$E(B-V)$}& \colhead{$T_{\mathrm{bb}}$} & \colhead{$L_{\rm X}$} \\
\colhead{} & \colhead{} & \colhead{(J2000)} & \colhead{(J2000)}& \colhead{(\arcsec)} & \colhead{(Mpc)} & \colhead{(mag)}&\colhead{(keV)} & \colhead{($10^{38}$ \ergs)} 
}
\startdata
2&M31&  0:42:43.288&$+$41:13:19.64&0.42 &0.78&0.42&0.15&0.10\\
3&M31&  0:42:47.868&$+$41:15:49.99&0.41 &0.78&0.50&0.13&0.05\\
4&M31&  0:42:52.509&$+$41:15:40.11&0.39 &0.78&0.49&0.06&7.8\\
13&M33& 1:33:35.927&$+$30:36:27.75&0.51 &0.82&0.036&0.15&0.04 \\
35&NGC\,2403&7:36:42.018&$+$65:36:51.75&0.39 &3.18&0.03&0.16&0.79\\
40&M81& 9:55:42.145&$+$69:03:36.37&0.36 &3.6&0.069&0.08&32\\
41&M81& 9:55:53.005&$+$69:05:20.15&0.49 &3.6&0.069&0.06&2.0\\
42&M81& 9:56:01.912&$+$68:58:59.28&0.44 &3.6&0.069&0.19&0.32\\
67&M51& 13:29:43.307&$+$47:11:34.91&0.39 &8&0.031&0.10&100\\
69&M51& 13:29:53.572&$+$47:11:26.46&0.37  &8&0.031&0.25&0.32\\
70&M51& 13:29:55.453&$+$47:11:43.49&0.41  &8&0.031&0.18&0.32\\
71&M51& 13:29:55.861&$+$47:11:44.69&0.41 &8&0.031&0.15&0.50\\
72&M83& 13:36:49.218&$-$29:53:02.99&0.46 &4.6&0.061&0.27&0.36\\
73&M83& 13:36:49.114&$-$29:52:58.50&0.42 &4.6&0.061&0.37&1.0\\
76&M83& 13:37:01.196&$-$29:54:49.15&0.41 &4.6&0.060&0.08&1.0\\
78&M83& 13:37:06.194&$-$29:52:31.71&0.39 &4.6&0.058&0.11&0.50\\
84&M101& 14:03:12.463&$+$54:17:54.09&0.34 &6.4&0.0076&0.16&0.25\\
87&M101& 14:03:18.971&$+$54:17:19.06&0.46 &6.4&0.0076&0.05&7.9\\
88&M101& 14:03:32.358&$+$54:21:02.58&0.37 &6.4&0.0075&0.12&26\\
89&M101& 14:03:33.328&$+$54:17:59.81&0.39 &6.4&0.0077&0.07&1.0\\
90&M101& 14:03:41.325&$+$54:19:04.02&0.39 &6.4&0.0076&0.14&5.0\\
95&NGC\,6946& 20:35:00.117&$+$60:09:08.09&0.37 &5.9&0.29&0.13&50\\
\enddata
\tablecomments{
Column (1): object ID in \citet{Zhou2019}.
Column (2): host galaxy.
Column (3): right ascension of the corrected X-ray position.
Column (4): declination of the corrected X-ray position.
Column (5): 90\% position error radius.
Column (6): luminosity distance to the host galaxy, quoted from \citet{Zhou2019} with references therein.
Column (7): Galactic extinction along the line of sight \citep{Schlafly2011}.
Column (8): Blackbody temperature quoted from \citet{Zhou2019}.
Column (9): X-ray luminosity quoted from \citet{Zhou2019}.
}
\end{deluxetable*}

\section{Sample, observations, and analysis}
\label{sec:sample}

We started with the sample of luminous and very soft X-ray sources in \citet{Zhou2019}.
These are point-like X-ray sources  with $L_{\rm X} \geq 10^{37}$~\ergs\ and  $T_{\rm bb} \leq 0.4$~keV found in Chandra observations of nearby galaxies, and are argued to contain candidates of compact objects with supercritical accretion viewed close to edge-on.
We then searched in the archive of HST, and found 22 X-ray sources in 8 galaxies with HST imaging observations available,  constituting the sample of this study (see Table~\ref{tab:sample}). 
In the following, we use the host galaxy name and the object ID in \citet{Zhou2019} to denote the source.
We note that there are two objects with $L_{\rm X} < 10^{37}$~\ergs, because \citet{Zhou2019} initially selected objects from \citet{She2017}, in which a different spectral model and/or distance was assumed and a slightly higher luminosity was inferred.
We still keep them to be consistent with \citet{Zhou2019}.

HST imaging observations using the Advanced Camera for Surveys (ACS) or Wide Field Camera 3 (WFC3) are used for optical counterpart search and photometry. 
The {\tt drc} files that have been corrected for geometric distortion and charge transfer inefficiency are used for analysis.
For each galaxy, we chose a Chandra Advanced CCD Imaging Spectrometer (ACIS) image to align with optical using a reference object shown in both. 
For galaxies with multiple ACIS observations available, the one with the target and reference objects located closest to the optical axis and with a relatively long exposure is used (typically in the range of 40-200 ks). 
We employed the {\tt chandra\_repro} tool in the CIAO package to create new level-2 event files.

\subsection{Astrometry and optical counterparts}

To enlarge the chance of identifying a unique optical counterpart of an X-ray source in nearby galaxies, one needs to align the Chandra X-ray image and HST optical image to improve the relative astrometry using at least one reference object shown in both. 

To increase the field of view of HST images and the chance of finding a reference object, we used the {\tt drizzle\_pac} tool in Python to combine partially overlapped HST images with the same instrument and the same filter to produce a mosaic image.  
For each galaxy, we selected a reference object that is close to our target object and exhibits both X-ray and optical emission for alignment, listed in Table~\ref{tab:ref}. 
Using more reference objects can further reduce the relative position error, but in general requires more HST images for a larger mosaic, which is often unavailable and may introduce additional uncertainties if the angular distance between the reference and target is large. 
In practice, a single nearby reference object is sufficient for our purpose. 
We note that M101-88 (also known as M101 ULX-1) has a known optical counterpart \citep{M101ref} and is used for alignment.
For M51-67 (M51 ULX-2), whose optical counterpart was also identified with a relatively large chance probability \citep{Terashima2006}, we used another reference object identified through refined astrometry \citep{M51ref}.

We ran {\tt wavdetect} to determine the X-ray source positions from a 0.3--8~keV image produced using the {\tt fluximage} tool, and {\tt source-extractor} for optical sources. 
The {\tt wcsmatch} tool in CIAO was used to register the X-ray coordinates to optical. 
The relative position uncertainty after alignment is a geometric sum of the statistical X-ray position errors of the reference and target sources, which can be estimated using an empirical function given the source counts and off-axis angle \citep[see Eq.~5 in ][]{2005ApJ...635..907H}. 
The statistical position uncertainty of optical sources in HST images is negligible compared with that of Chandra.

\begin{deluxetable}{ccccc}
\tablecaption{Chandra observations and reference objects used for relative astrometry}
\label{tab:ref}
\tablehead{\colhead{Host } &\colhead{ObsID}  & \colhead{Ref.\ R.A.} & \colhead{Ref.\ Decl.}  & \colhead{Refs.}}
\startdata
M31&309& 00:42:59.875 &$+$41:16:05.67 & (1) \\
M33&1730& 01:33:41.898 &$+$30:38:48.79 & (2) \\
M51&13814& 13:30:06.450 &$+$47:08:34.70 & (3) \\
M81&735& 09:55:32.956 &$+$69:00:33.75 & (4) \\
M83&793& 13:37:19.801 &$-$29:53:48.72 & (5) \\
M101&934& 14:03:32.362 &$+$54:21:02.72 & (6) \\
NGC 2403&201& 07:36:25.559 &$+$65:35:39.95 & (7) \\
NGC 6946&1043& 20:34:56.270 &$+$60:09:07.31 & N.A. \\
\enddata
\tablerefs{
(1) \citet{M31ref}, 
(2) \citet{M33ref}, 
(3) \citet{M51ref}, 
(4) \citet{M81ref}, 
(5) \citet{M83ref}, 
(6) \citet{M101ref}, 
(7) \citet{NGC2403ref}.}
\end{deluxetable}

Chandra has a 90\% absolute position uncertainty of 0\farcs85 after reprocessing\footnote{\url{https://cxc.harvard.edu/cal/ASPECT/celmon/}}. 
After alignment with HST images, the relative position uncertainty calculated using the above recipe can be reduced to about 0\farcs4. 
The properties of the sample objects are also listed in Table~\ref{tab:sample}, including the host galaxy, corrected X-ray position and error radius, distance, foreground extinction, and X-ray luminosity and temperature. 

\begin{figure*}
\includegraphics[width=0.245\linewidth]{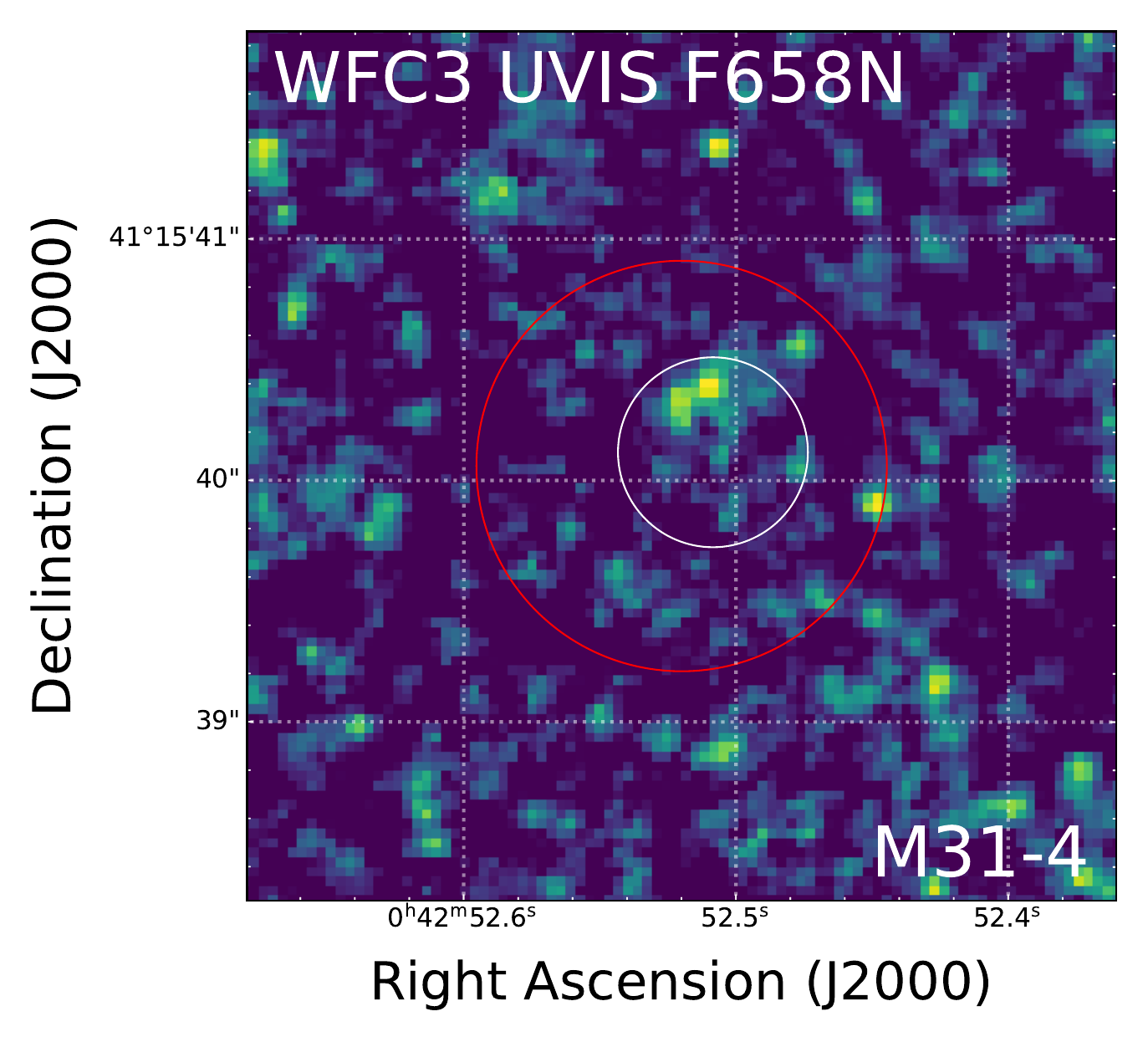}
\includegraphics[width=0.245\linewidth]{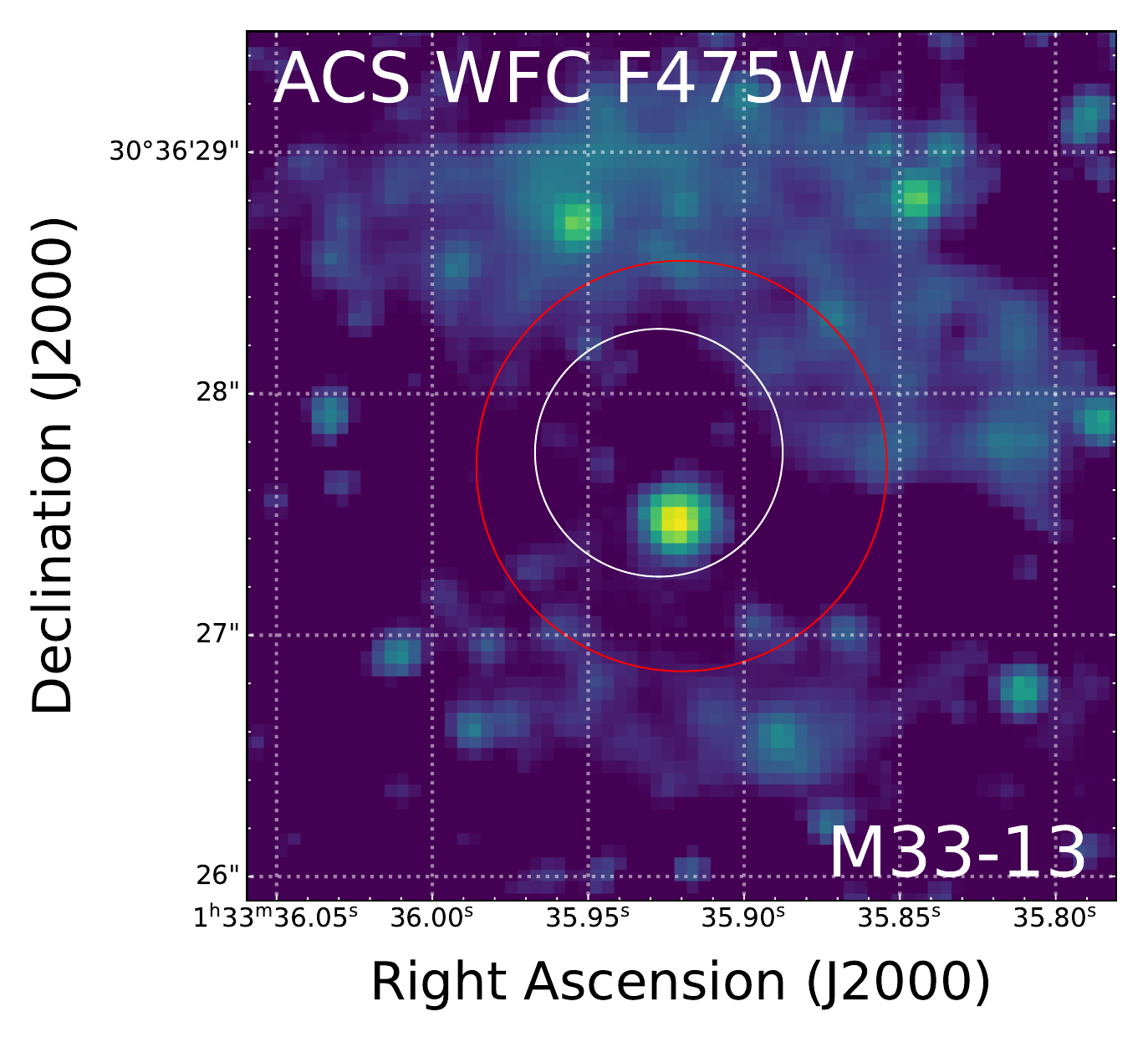}
\includegraphics[width=0.245\linewidth]{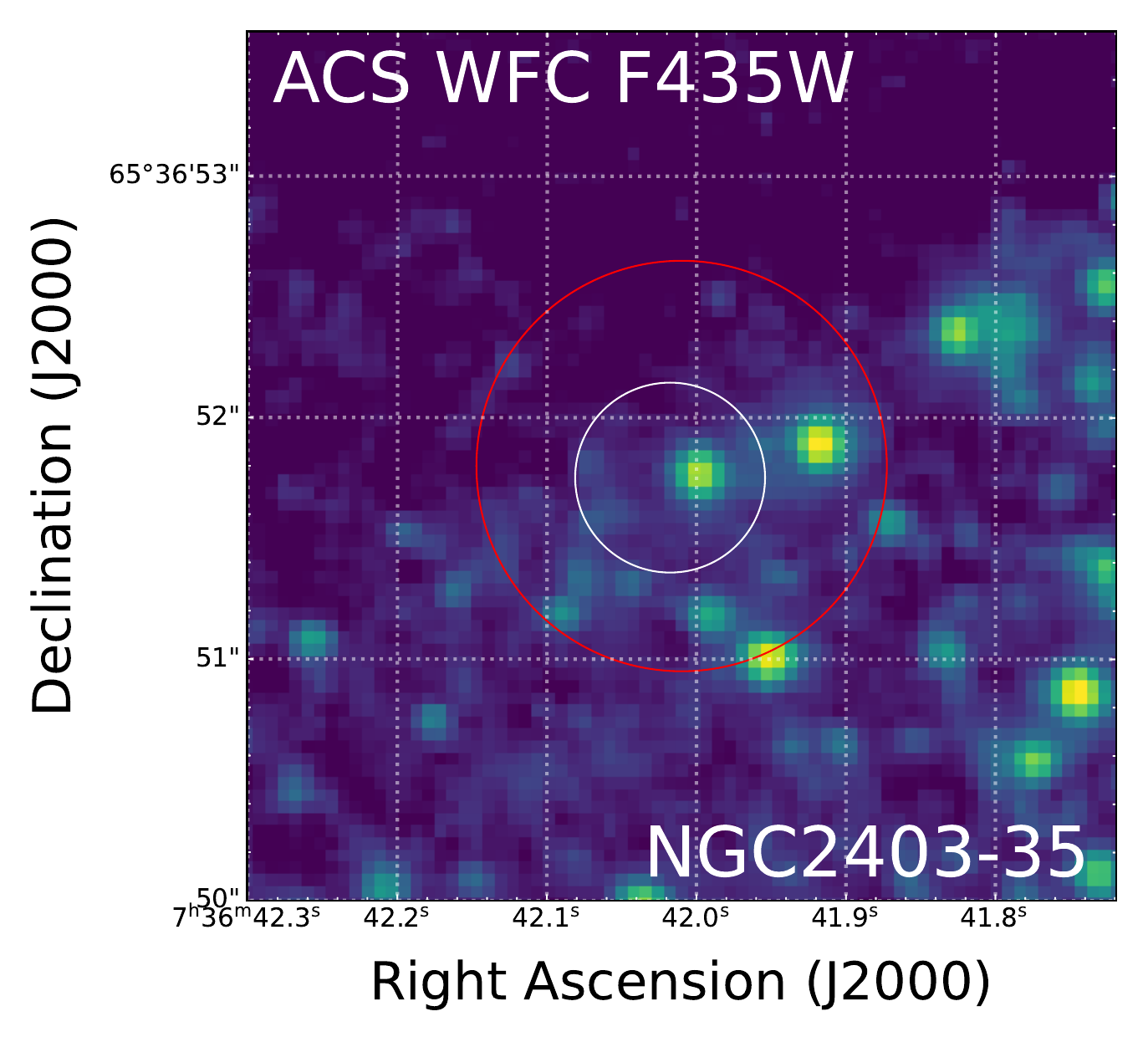}
\includegraphics[width=0.245\linewidth]{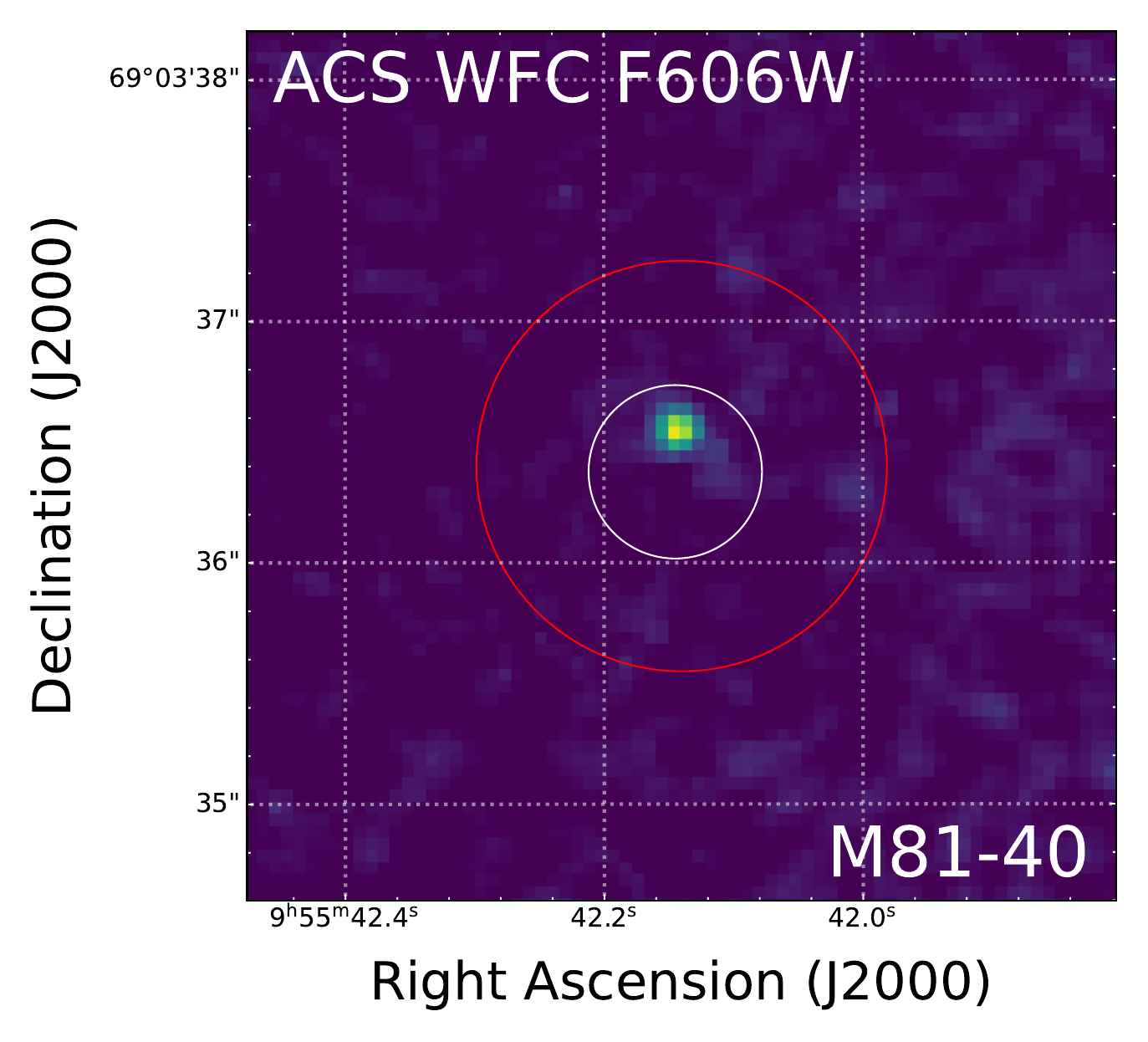}
\includegraphics[width=0.245\linewidth]{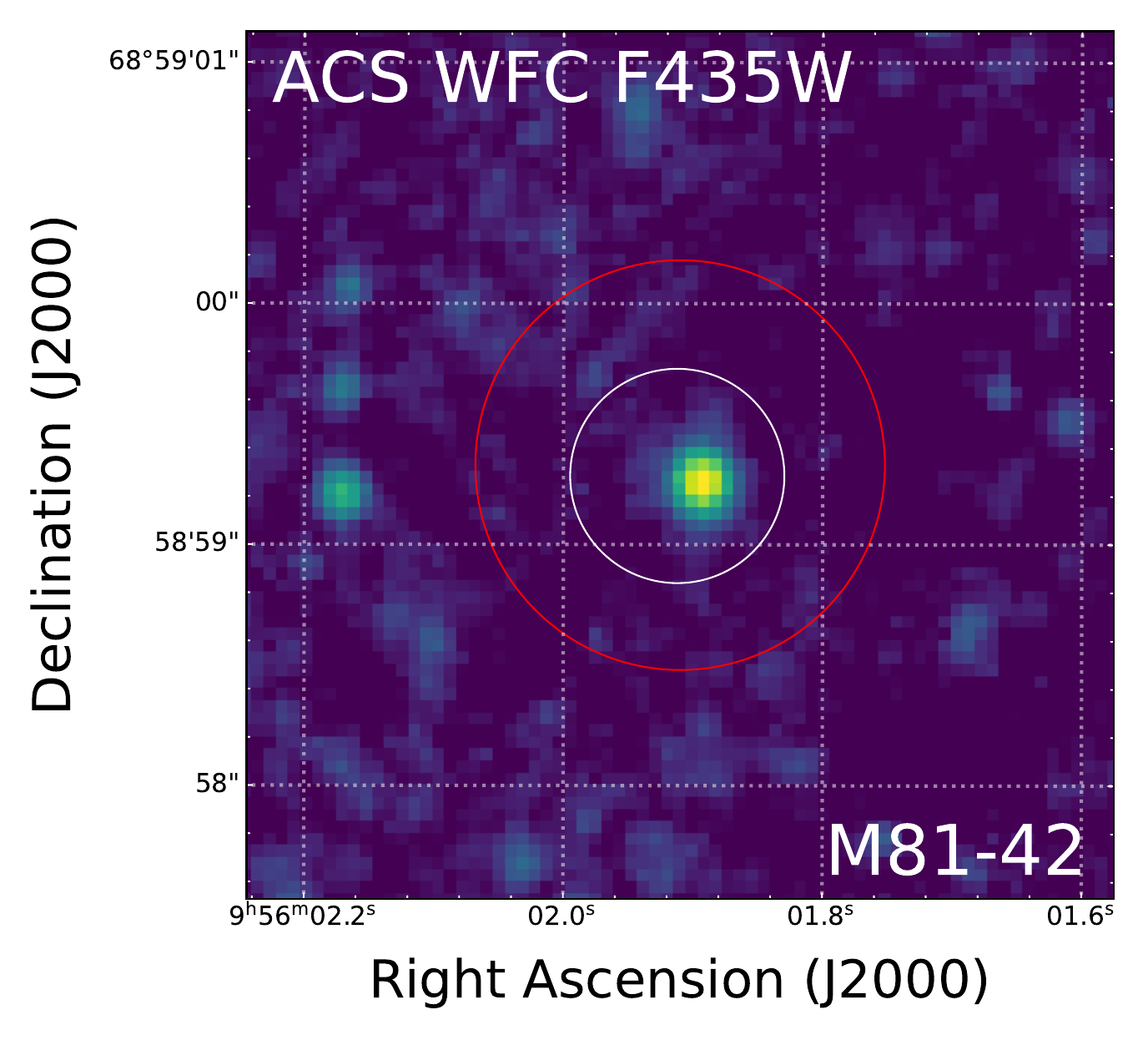}
\includegraphics[width=0.245\linewidth]{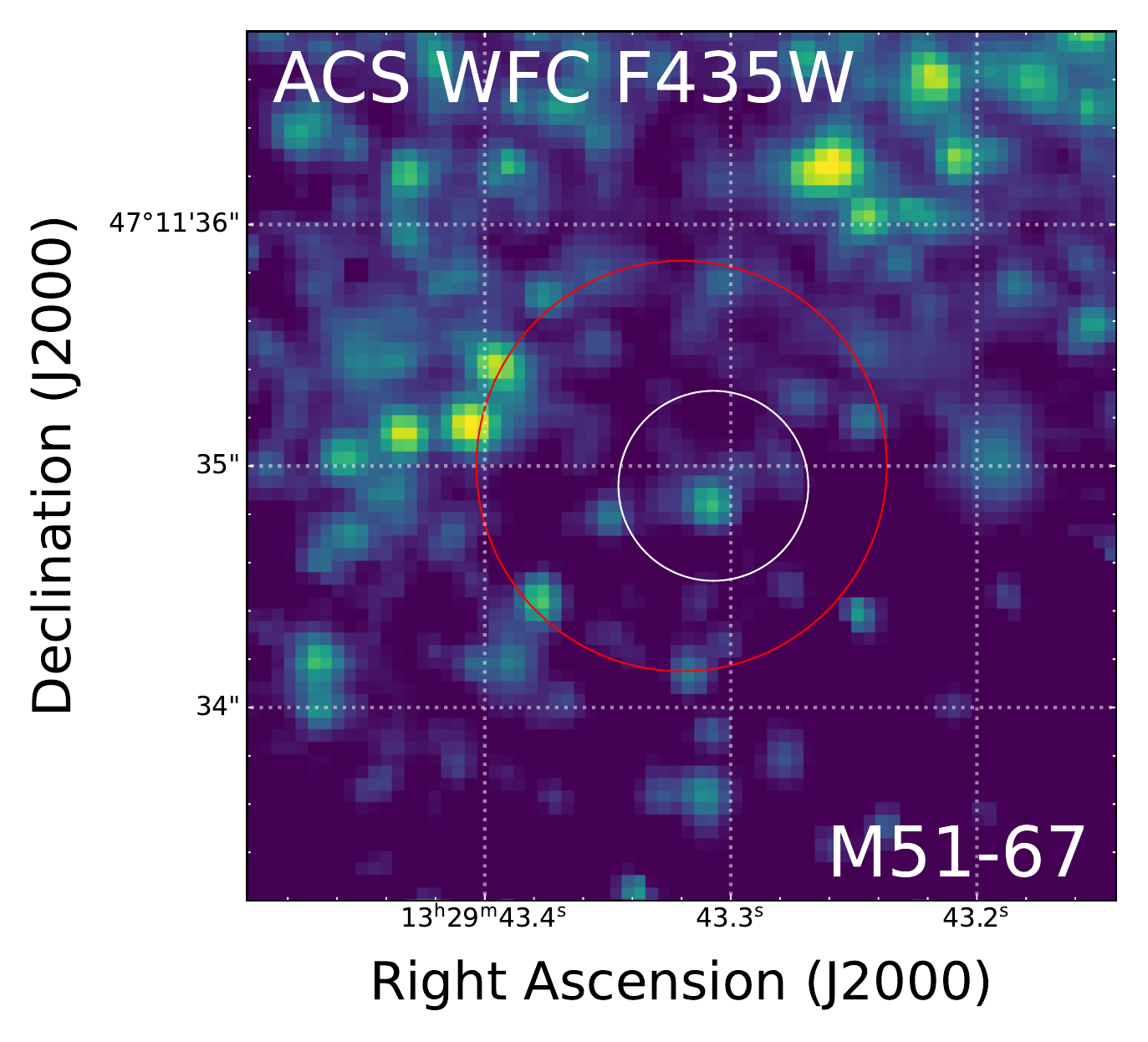}
\includegraphics[width=0.245\linewidth]{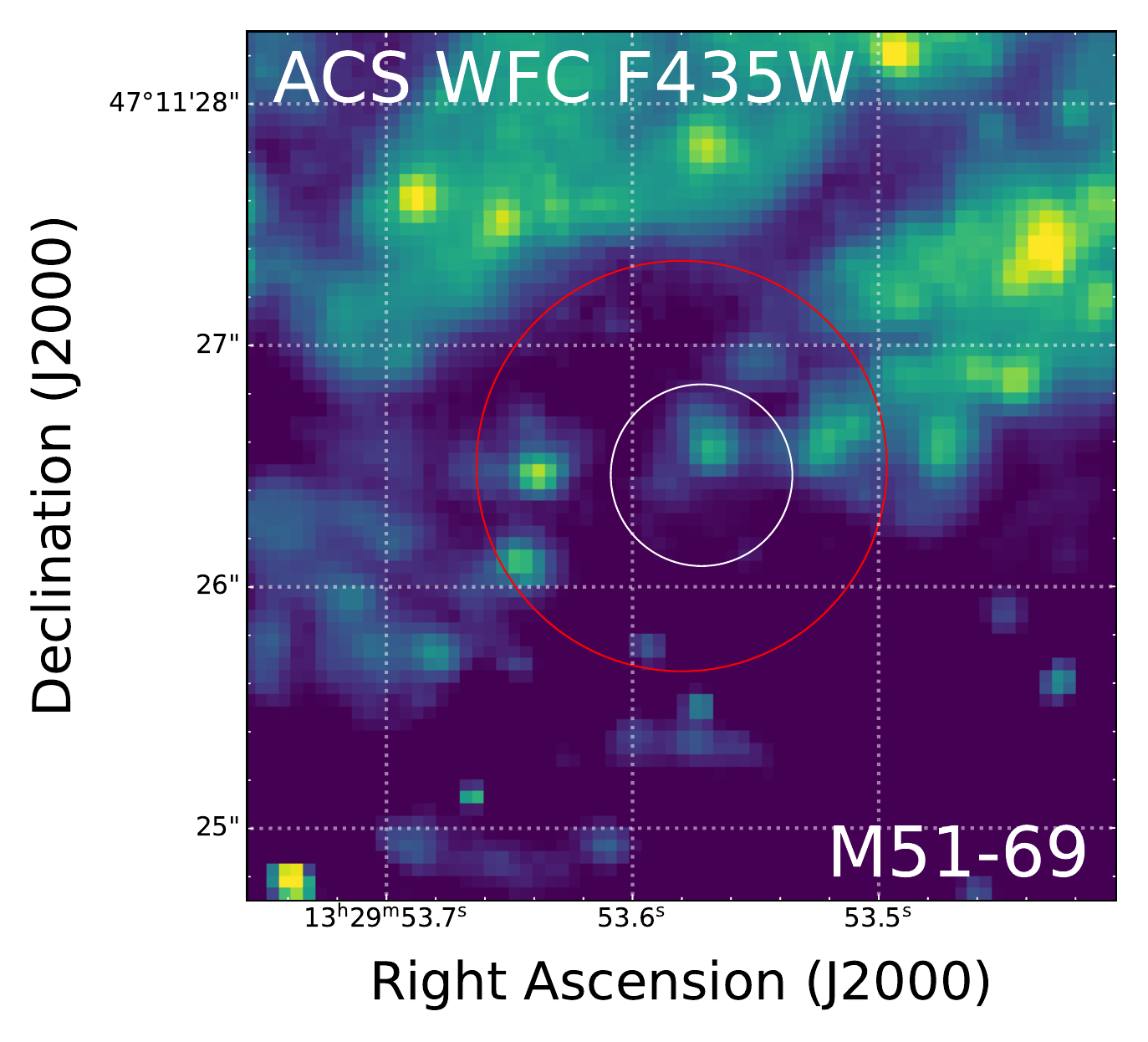}
\includegraphics[width=0.245\linewidth]{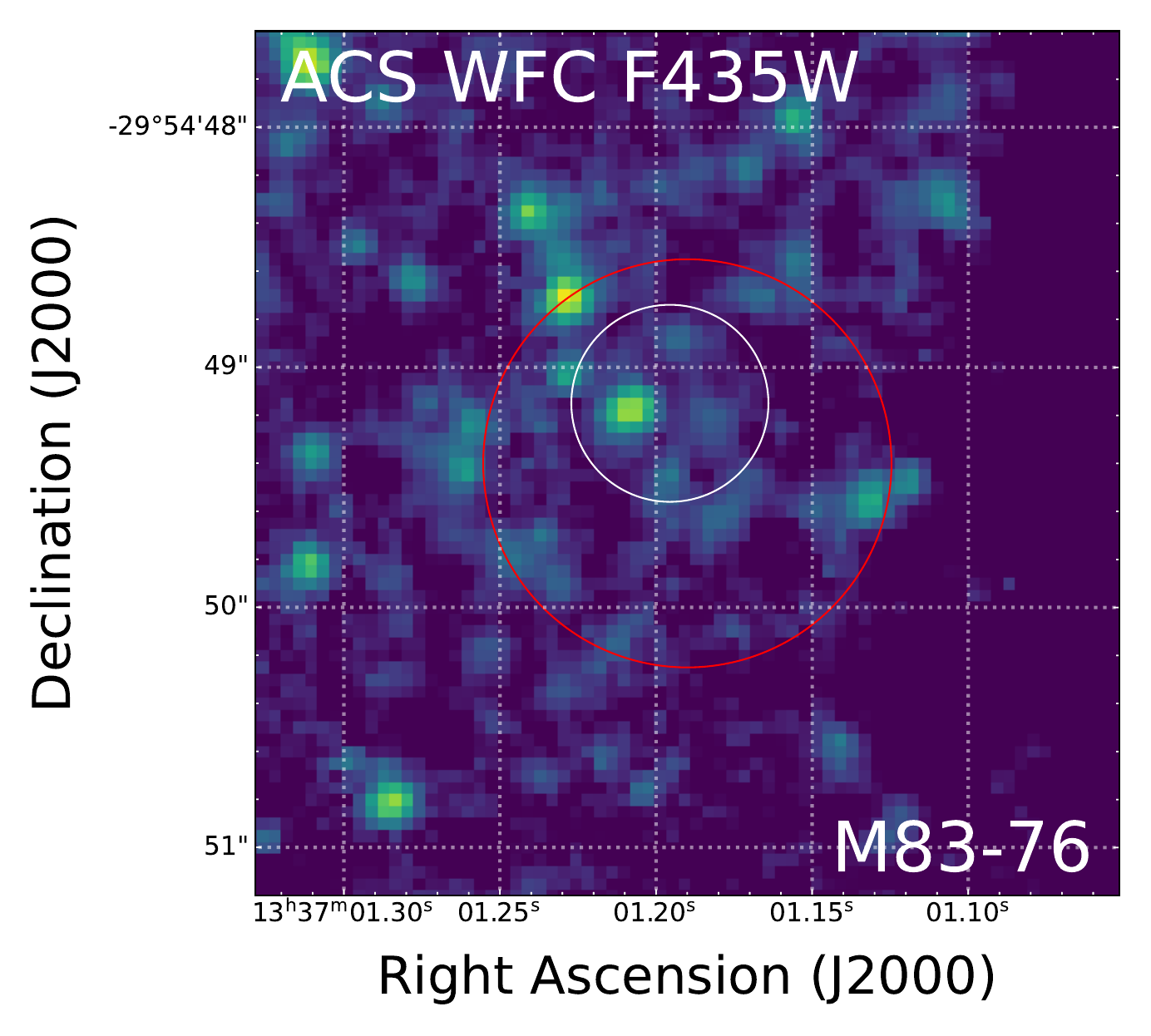}
\includegraphics[width=0.245\linewidth]{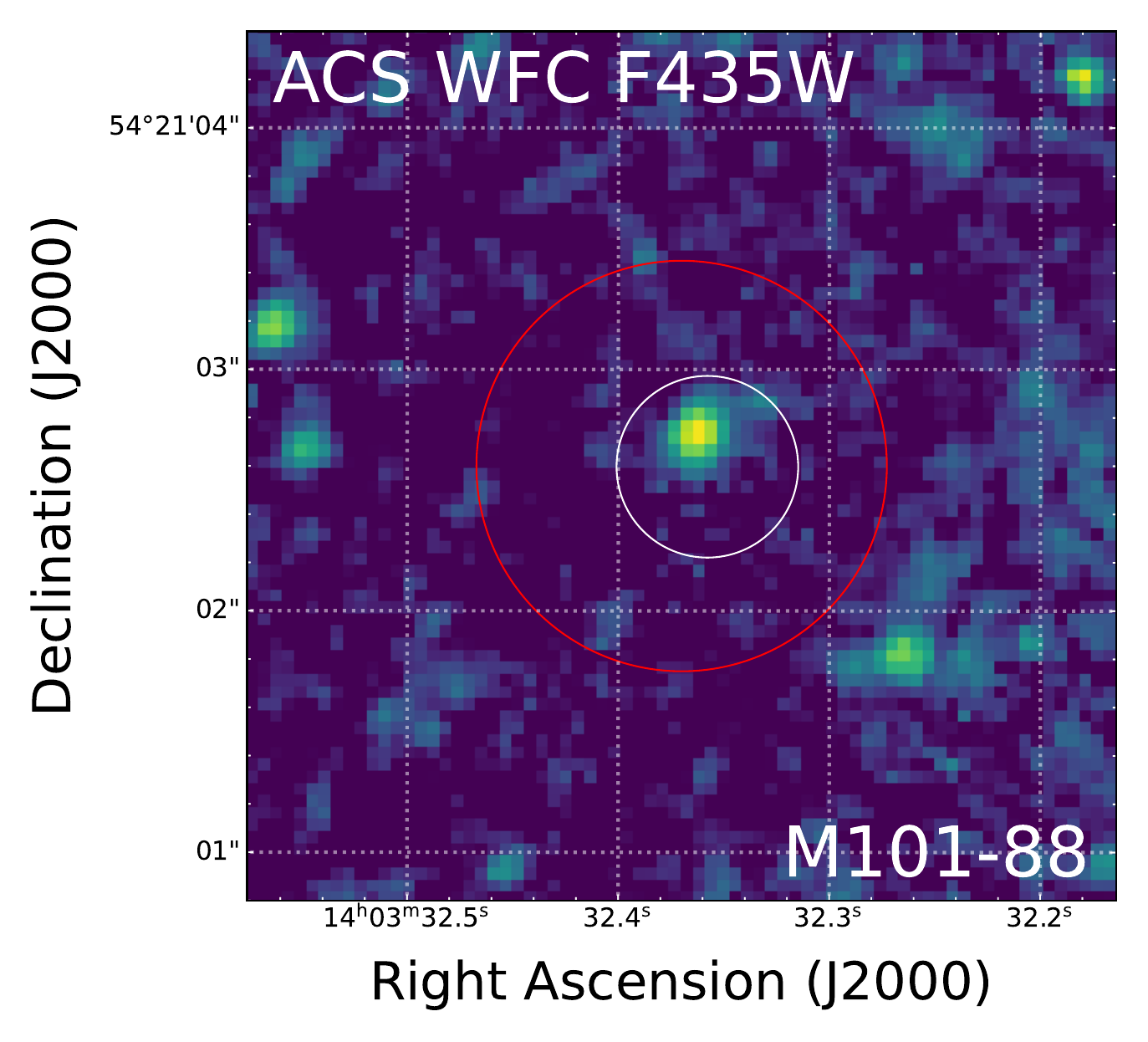}
\includegraphics[width=0.245\linewidth]{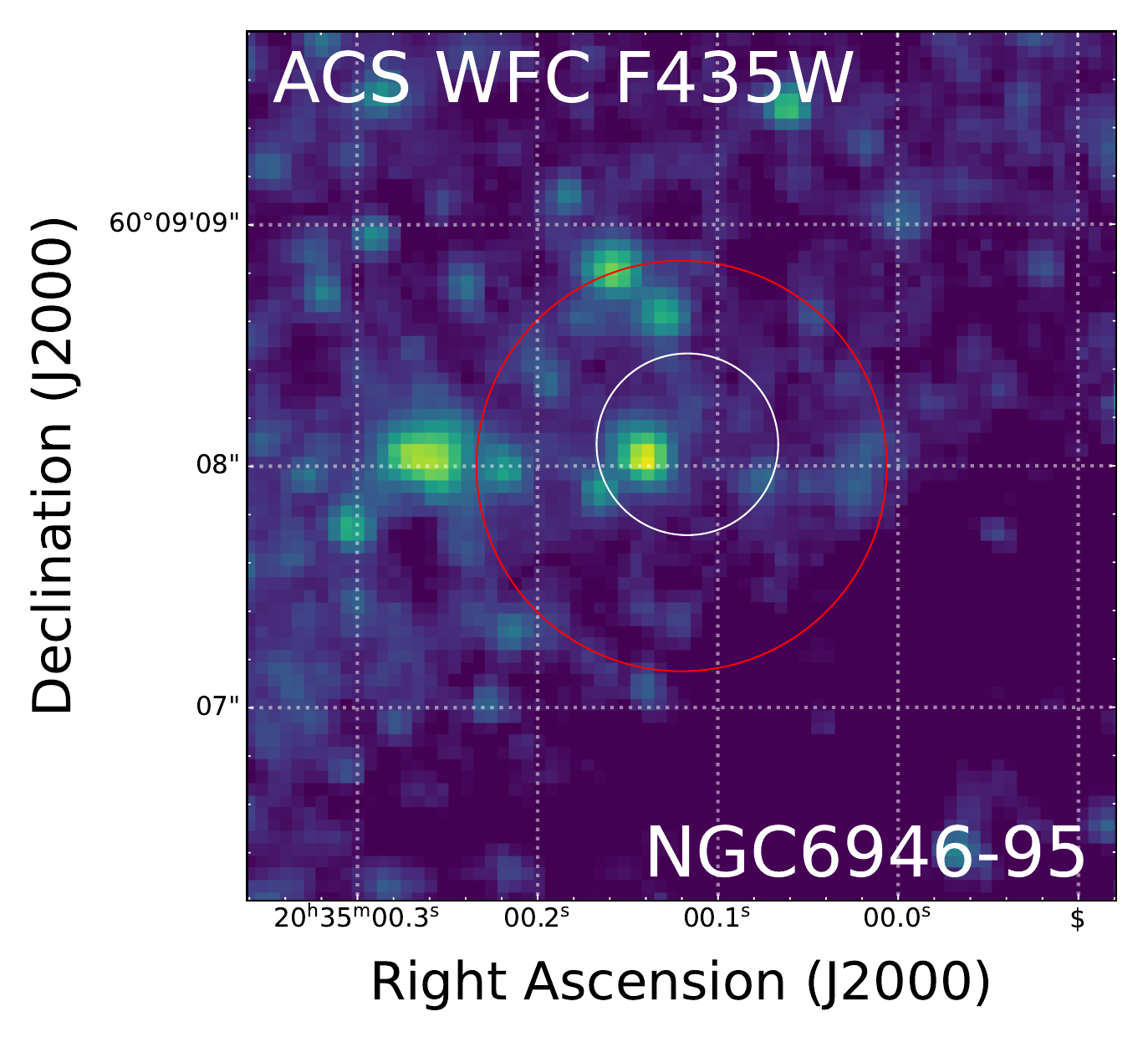}
\caption{HST images around the X-ray sources identified with an optical counterpart. The red circle indicates the absolute Chandra position uncertainty, and the white circle indicates the corrected position uncertainty after aligning the Chandra and HST images. The instrument setup, host galaxy, and source ID are displayed for each target. We choose an F435W image for display as this is the most commonly used filter; for M81-40, no F435W image is available and an F606W image is used instead; for M31-4, a WFC3 narrow-band image is used because the point-like source and the extended feature cannot be resolved in broadband ACS images.}
\label{fig:cp}
\end{figure*}

\begin{figure*}
\includegraphics[width=0.245\linewidth]{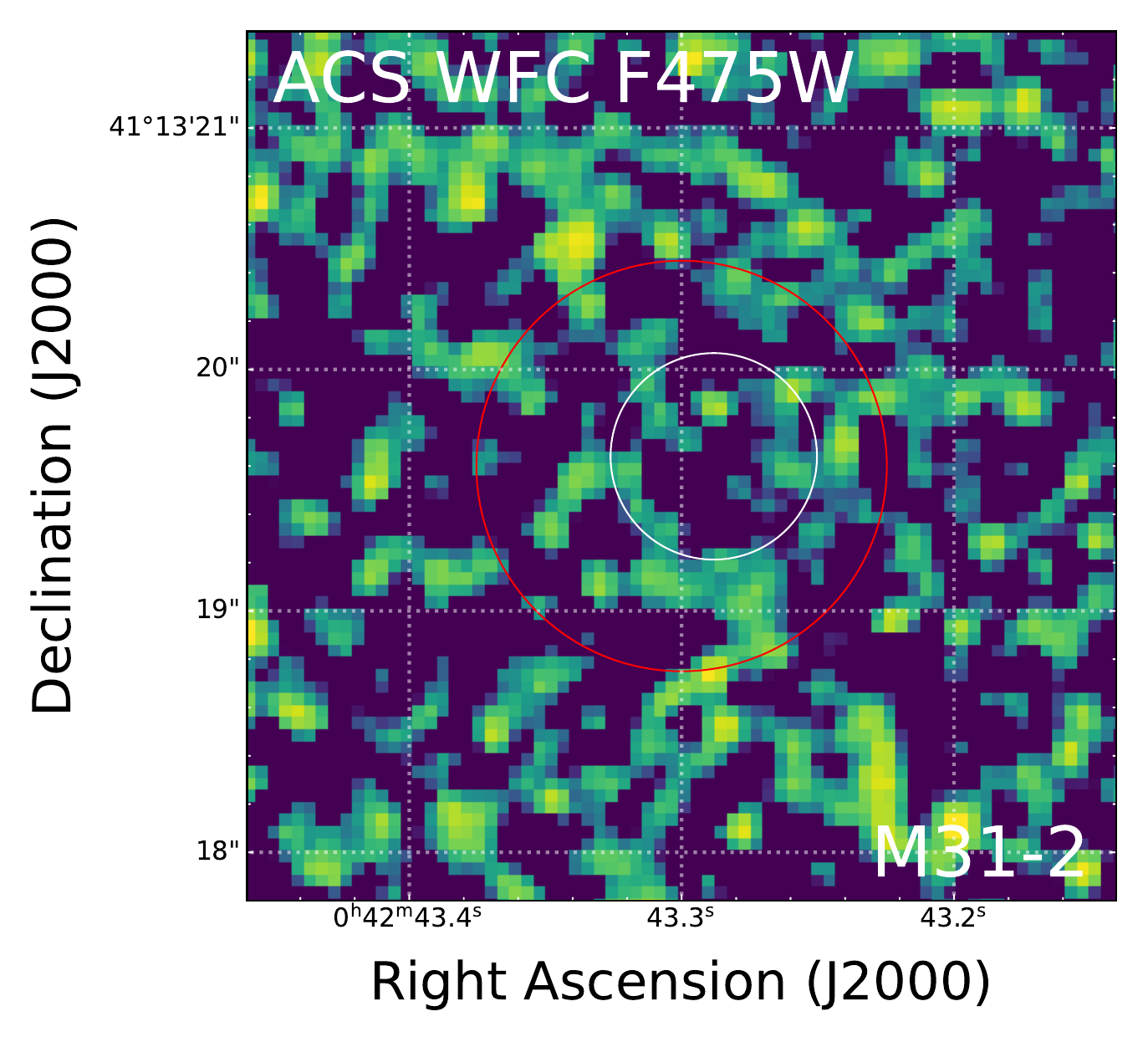}
\includegraphics[width=0.245\linewidth]{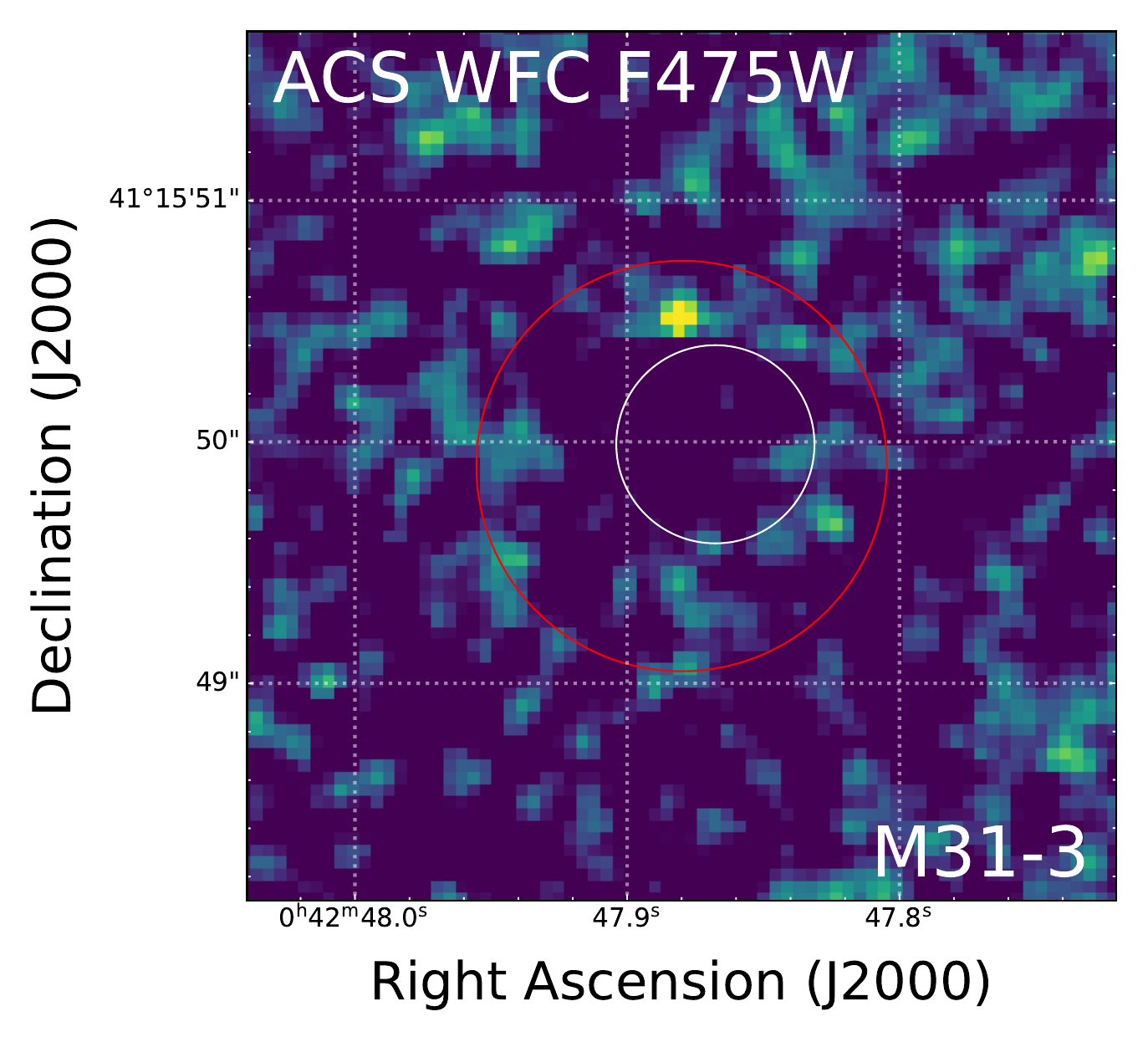}
\includegraphics[width=0.245\linewidth]{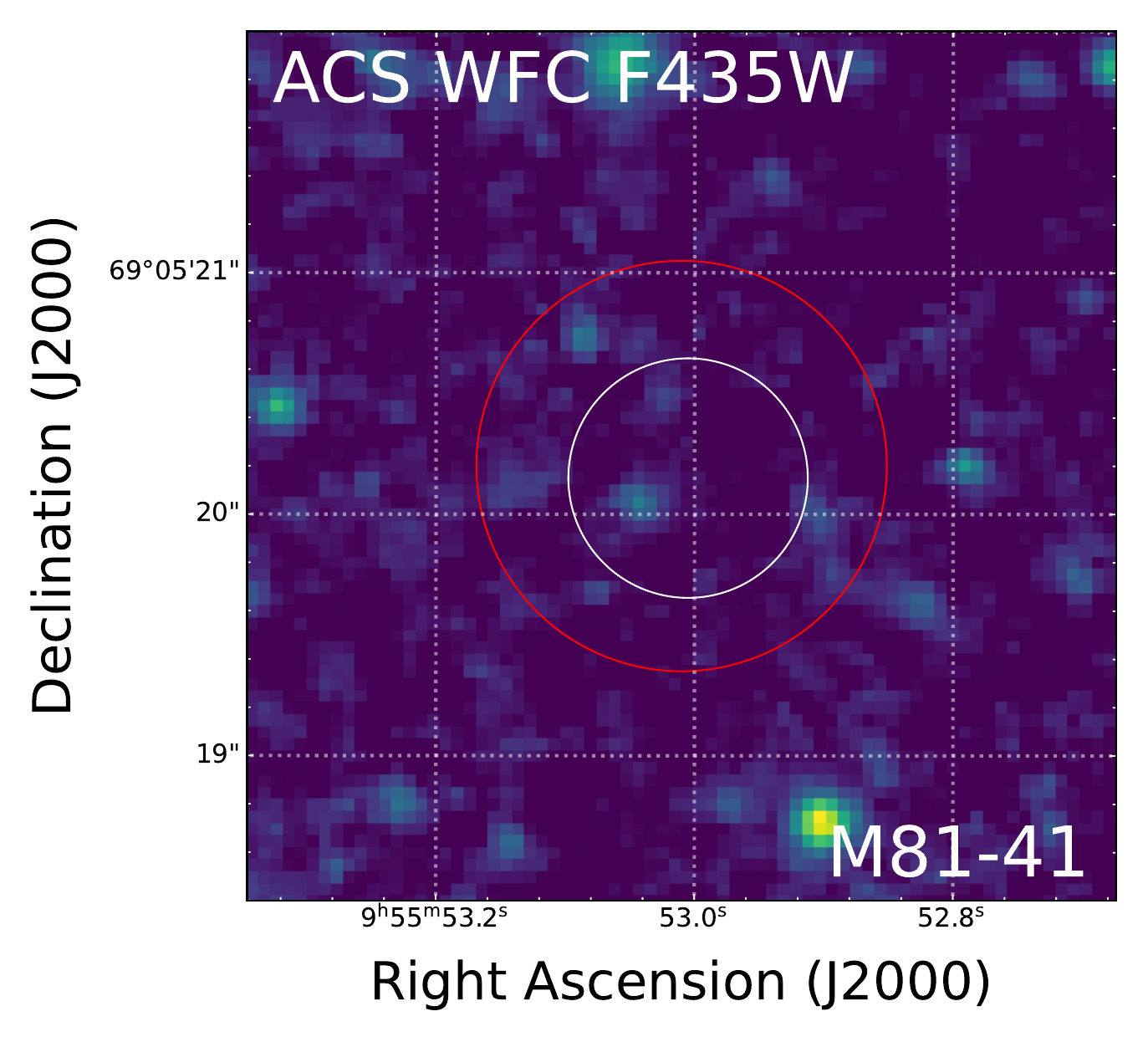}
\includegraphics[width=0.245\linewidth]{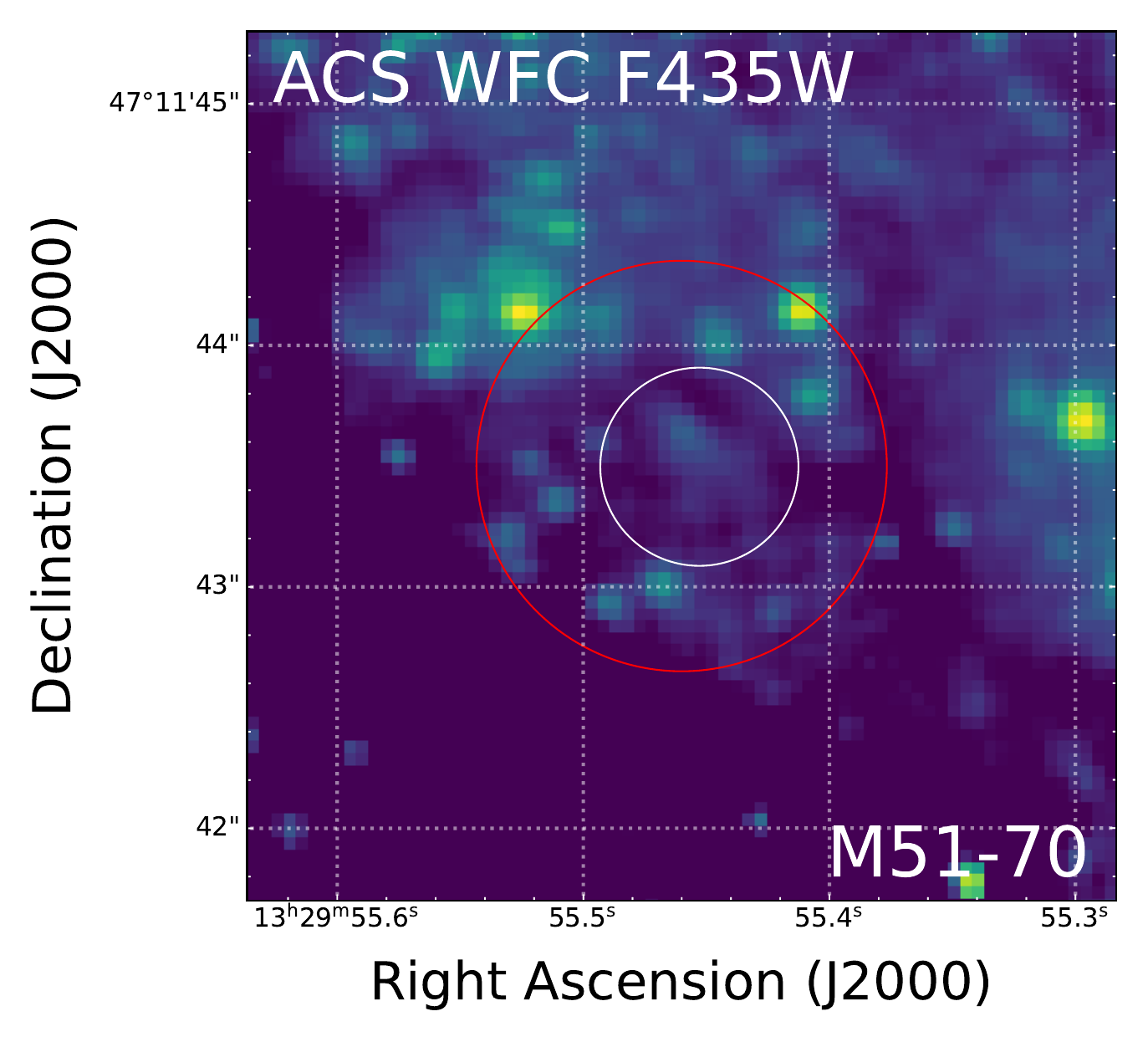}
\includegraphics[width=0.245\linewidth]{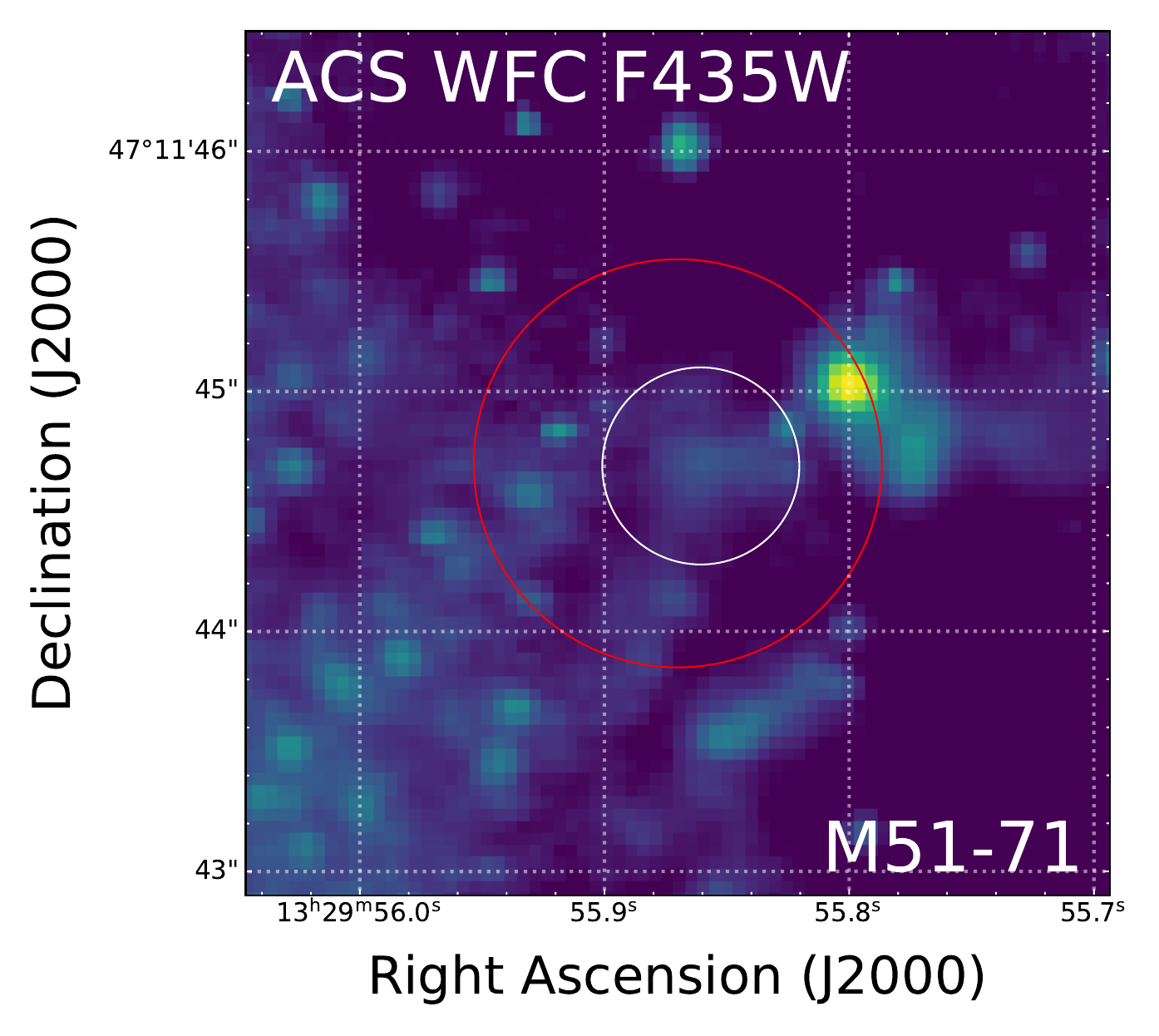}
\includegraphics[width=0.245\linewidth]{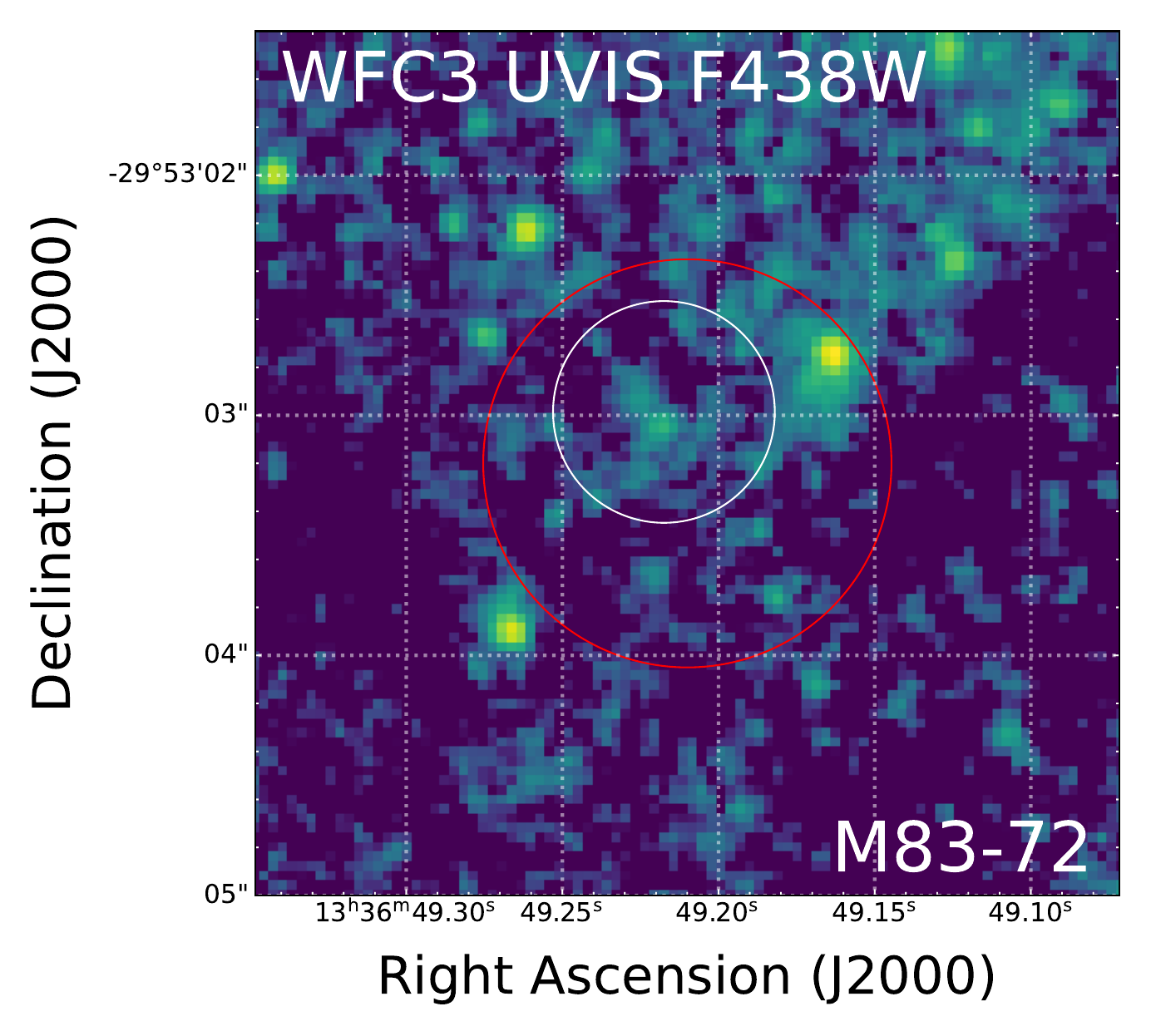}
\includegraphics[width=0.245\linewidth]{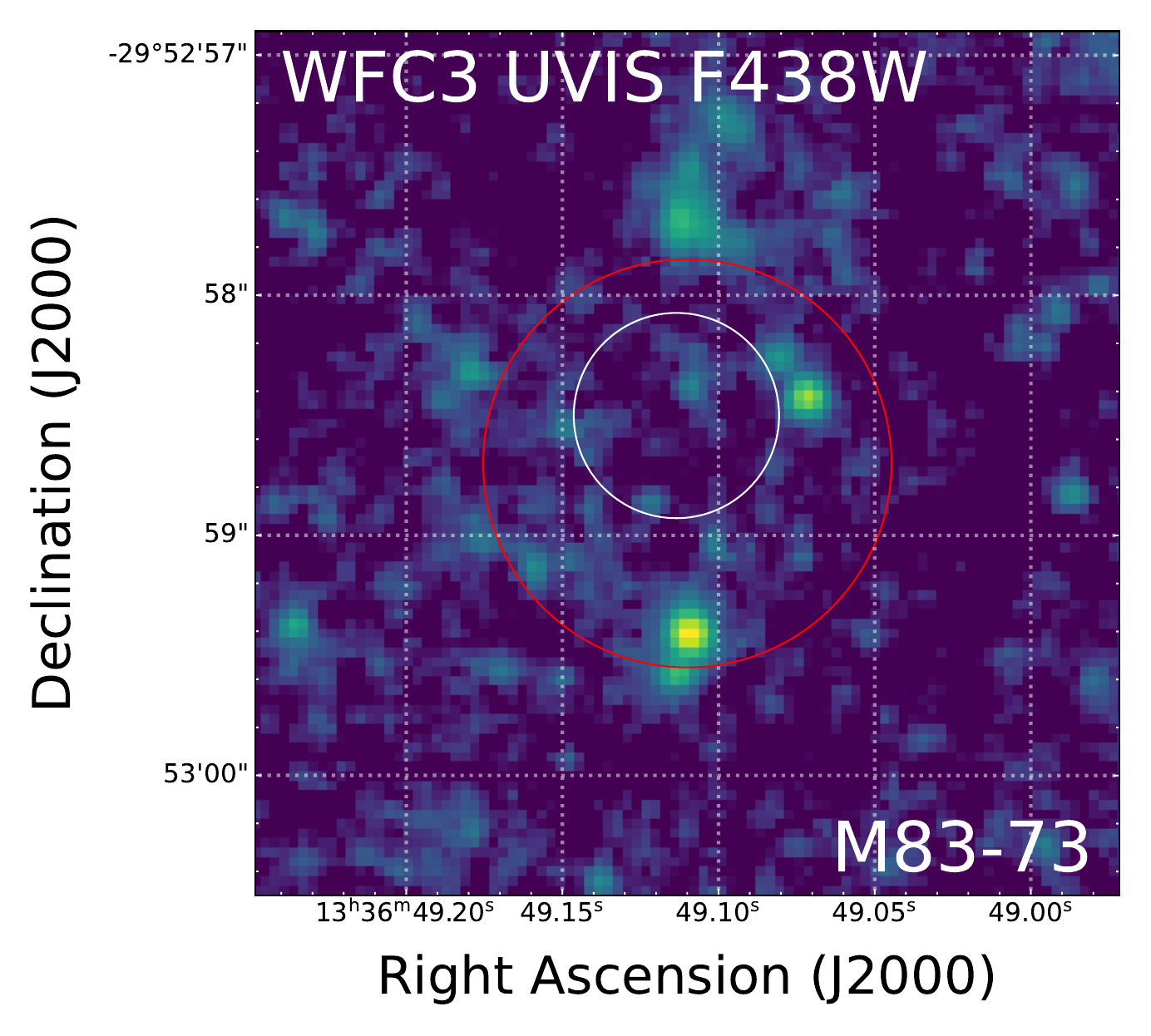}
\includegraphics[width=0.245\linewidth]{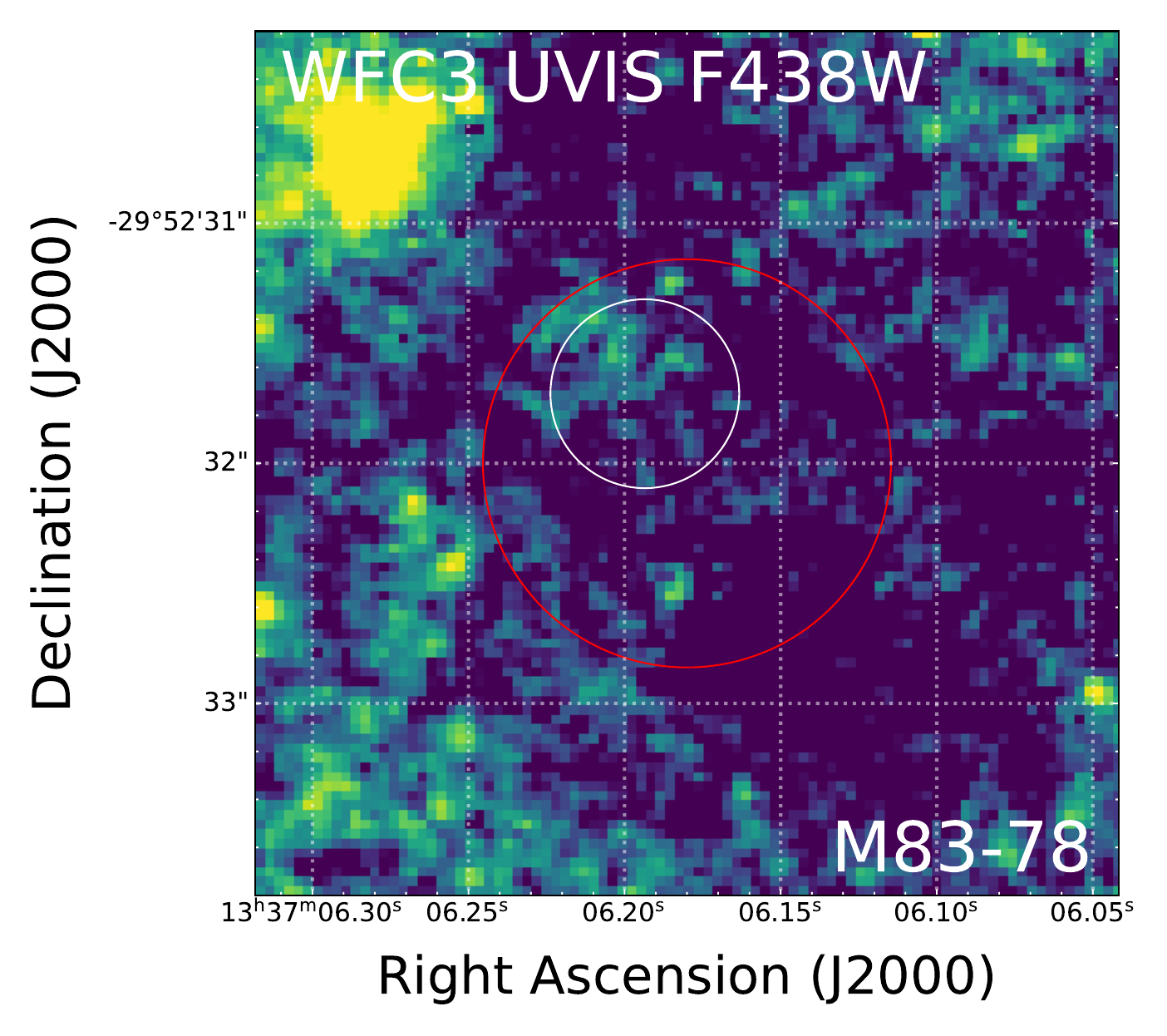}
\includegraphics[width=0.245\linewidth]{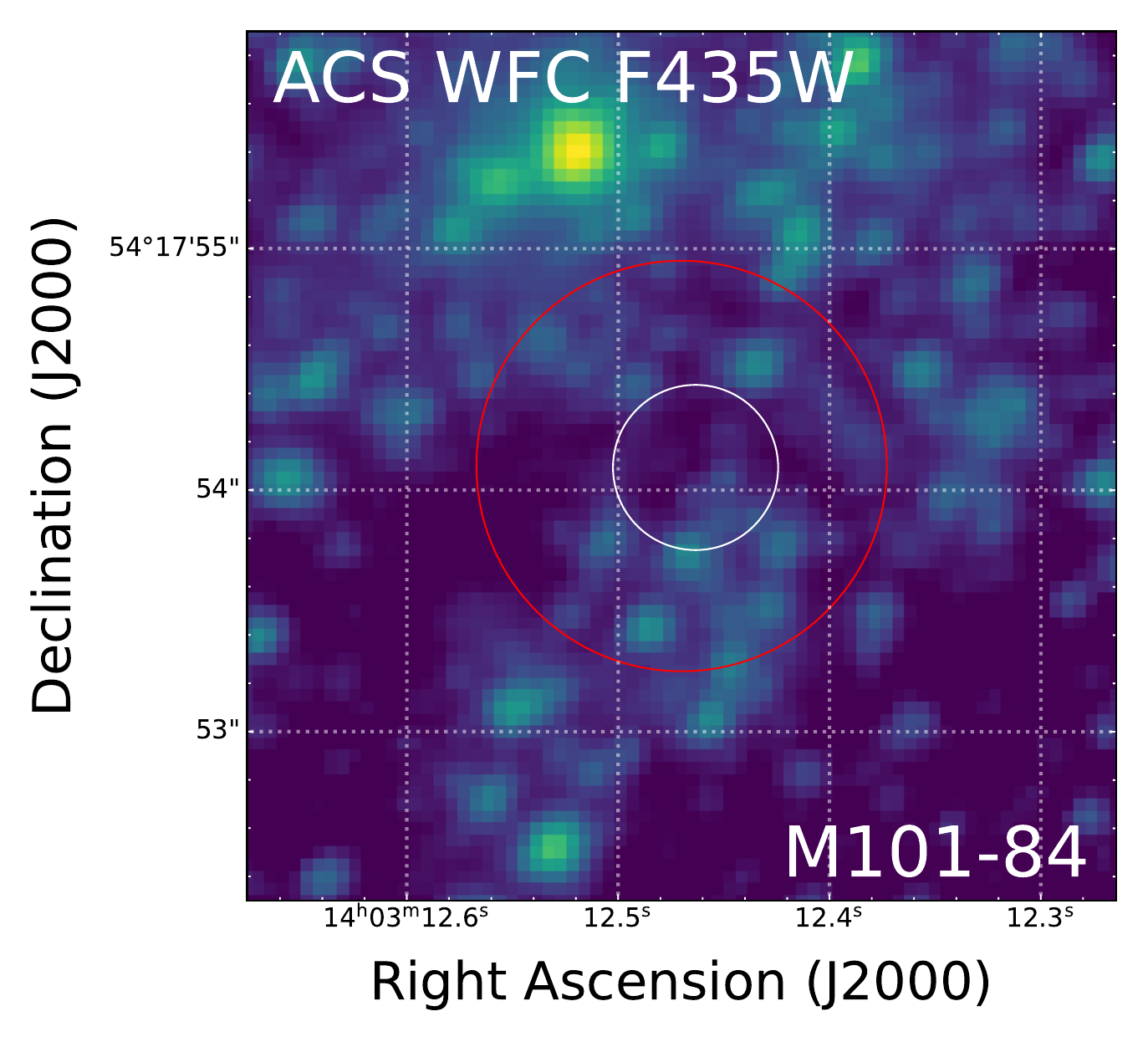}
\includegraphics[width=0.245\linewidth]{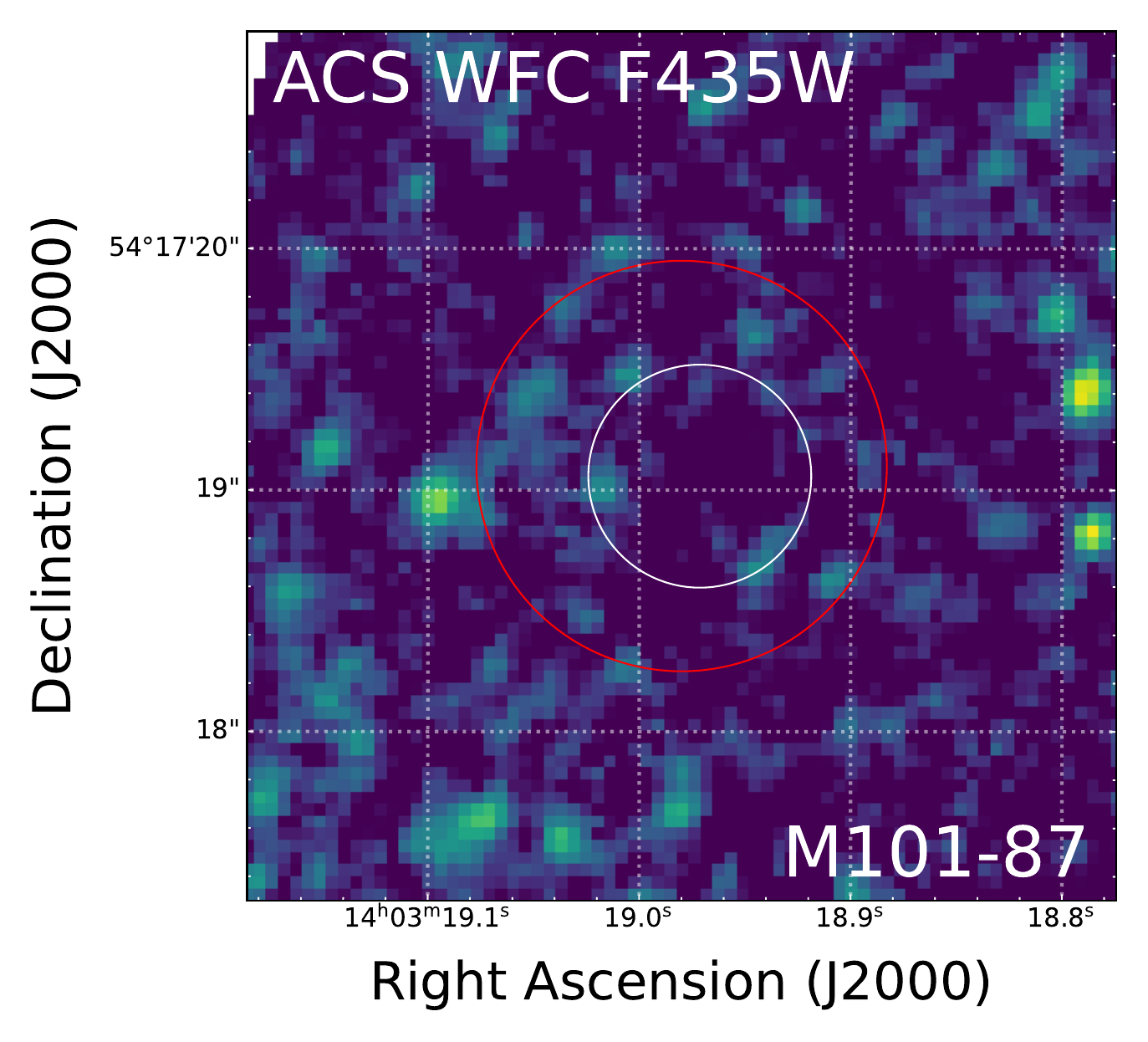}
\includegraphics[width=0.245\linewidth]{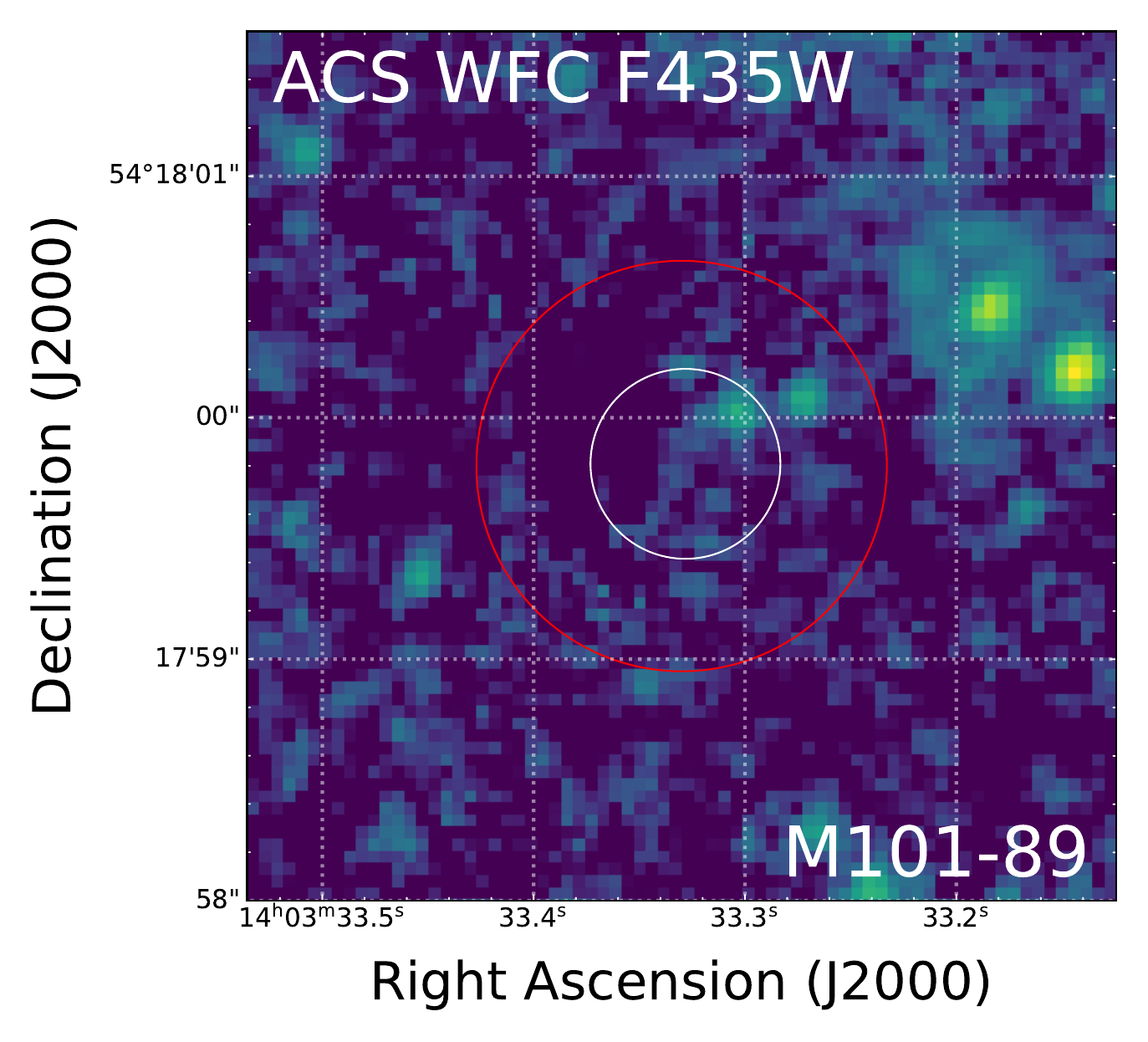}
\includegraphics[width=0.245\linewidth]{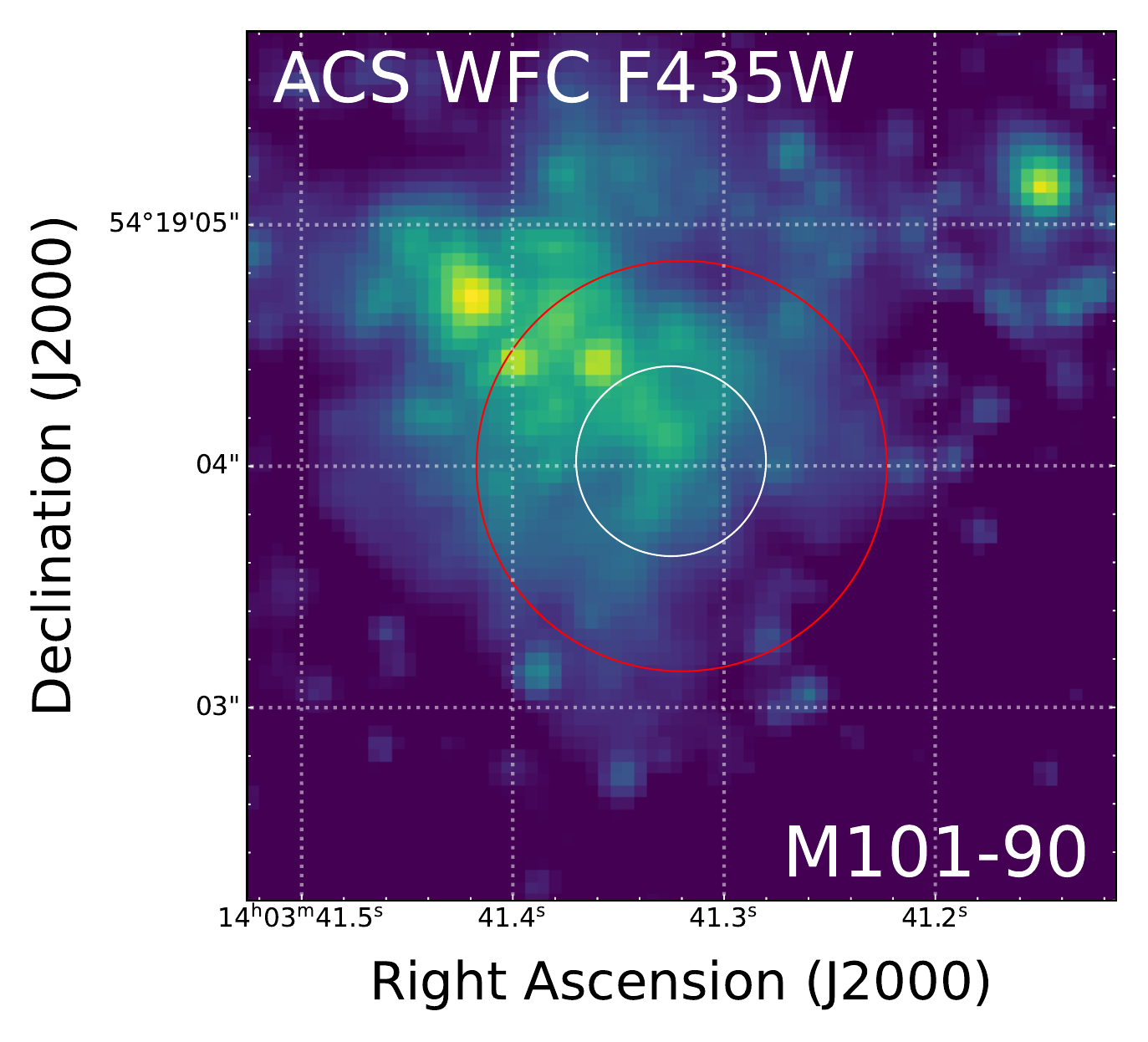}
\caption{HST images around the X-ray sources that no optical counterpart is identified. The circles and texts have the same meaning as in Figure~\ref{fig:cp}.}
\label{fig:cp_no}
\end{figure*}

The HST images around the X-ray sources are displayed in Figure~\ref{fig:cp} \& \ref{fig:cp_no}, respectively, for those with and without an identification of an optical counterpart. 
The absolute and relative position error circles are displayed in the figures. 
In this work, we only consider optical sources with $m < 24$ within the corrected 90\% error circle as potential optical counterparts, and found 10 such objects (Figure~\ref{fig:cp}).
The error radius is in the range of 0\farcs35--0\farcs50. 
The empirical threshold magnitude is adopted from \citet{Tao2011} for the ULX optical counterparts.
All of them are consistent with a point-like object in the HST image.
We note that for M31-4, the optical emission within the error circle consists of a point-like object (toward the northernmost region) and an extended component in the WFC3 UVIS F658N image (the two components cannot be resolved in ACS images).
We tentatively regard the point-like object as the optical counterpart. 
While for the other 12 targets (Figure~\ref{fig:cp_no}), no point-like optical counterparts can be identified in the image. 
Photometry is performed on the brightest emission region in the error circle to get a 2$\sigma$ upper limit on the optical flux. 

\begin{figure*}
\centering
\includegraphics[width=0.33\linewidth]{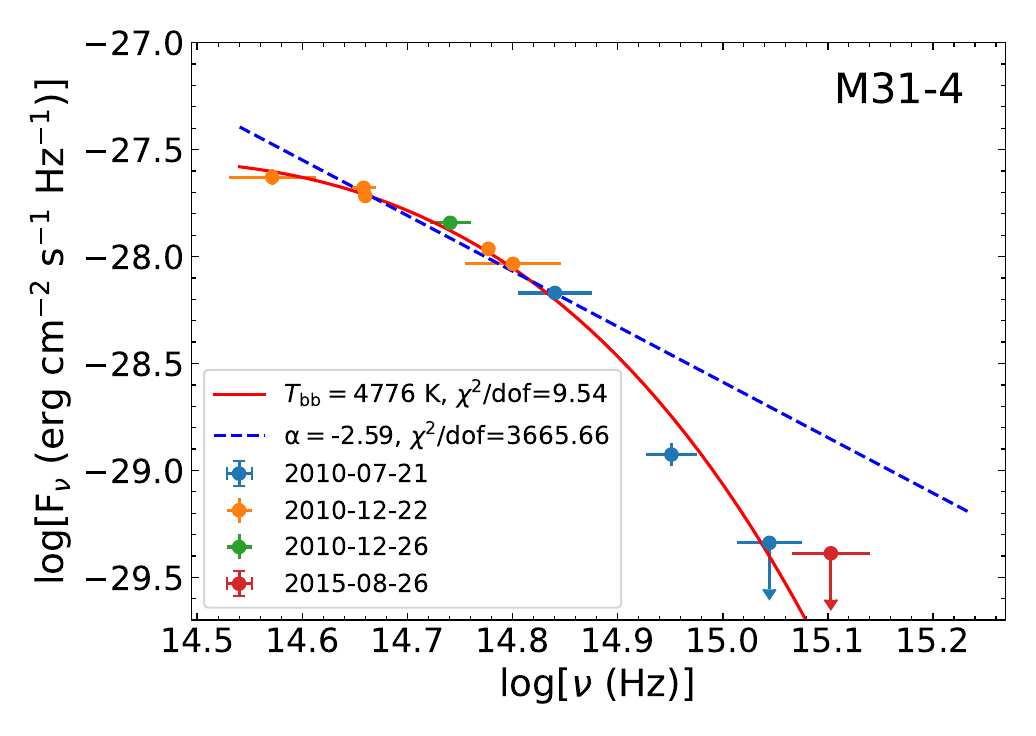}
\includegraphics[width=0.33\linewidth]{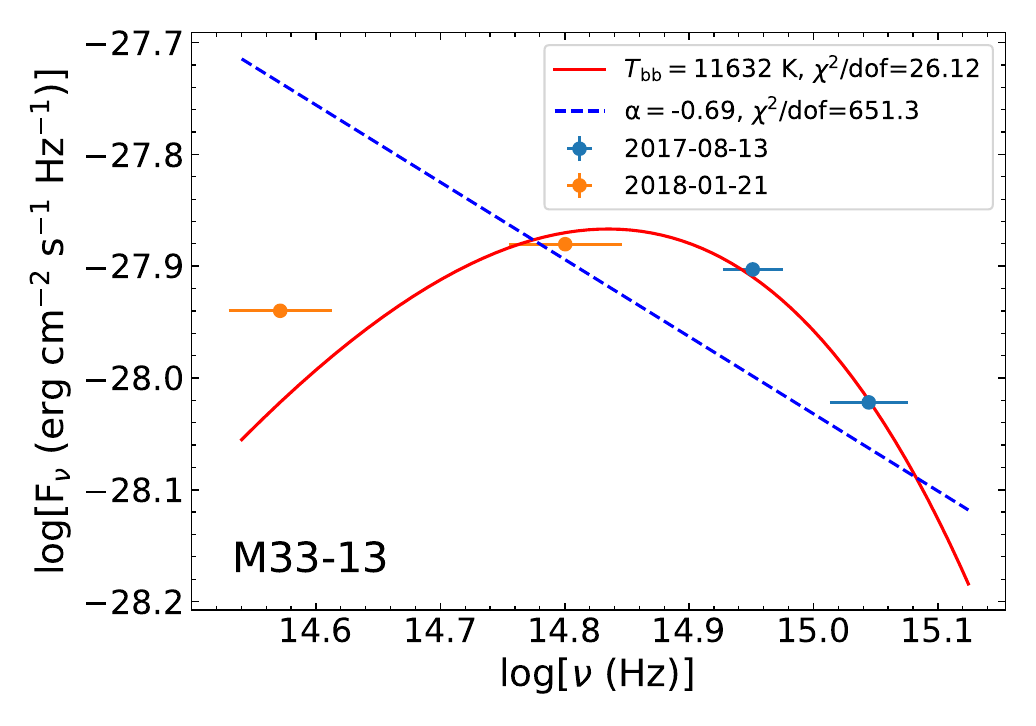}
\includegraphics[width=0.33\linewidth]{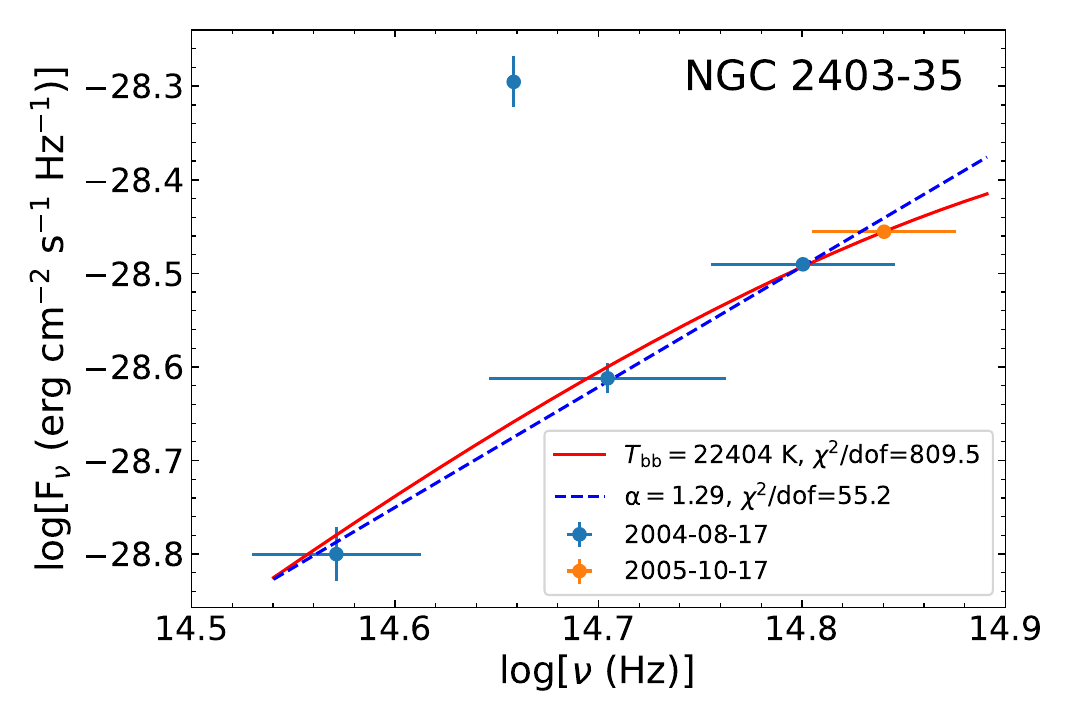}
\includegraphics[width=0.33\linewidth]{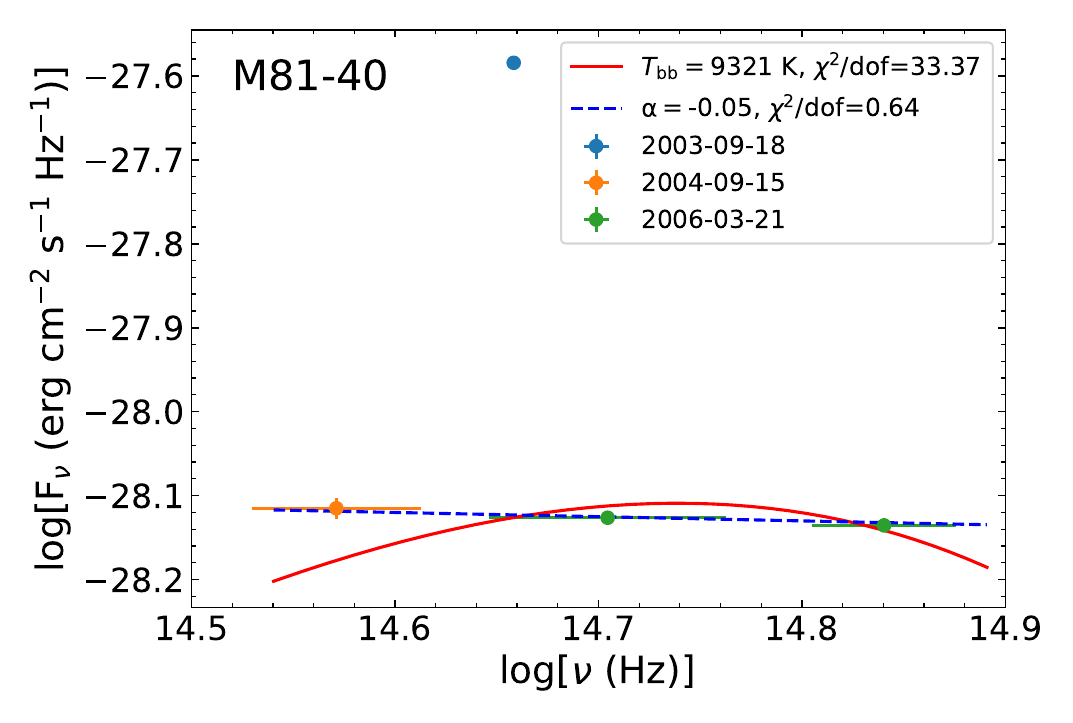}
\includegraphics[width=0.33\linewidth]{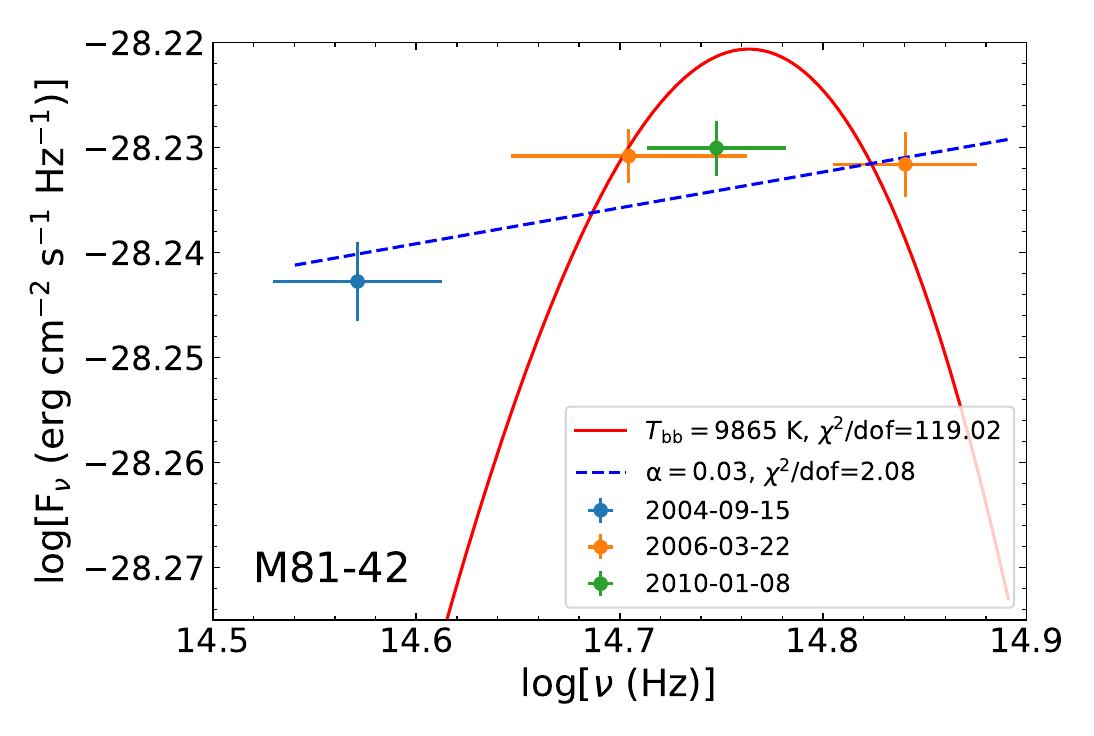}
\includegraphics[width=0.33\linewidth]{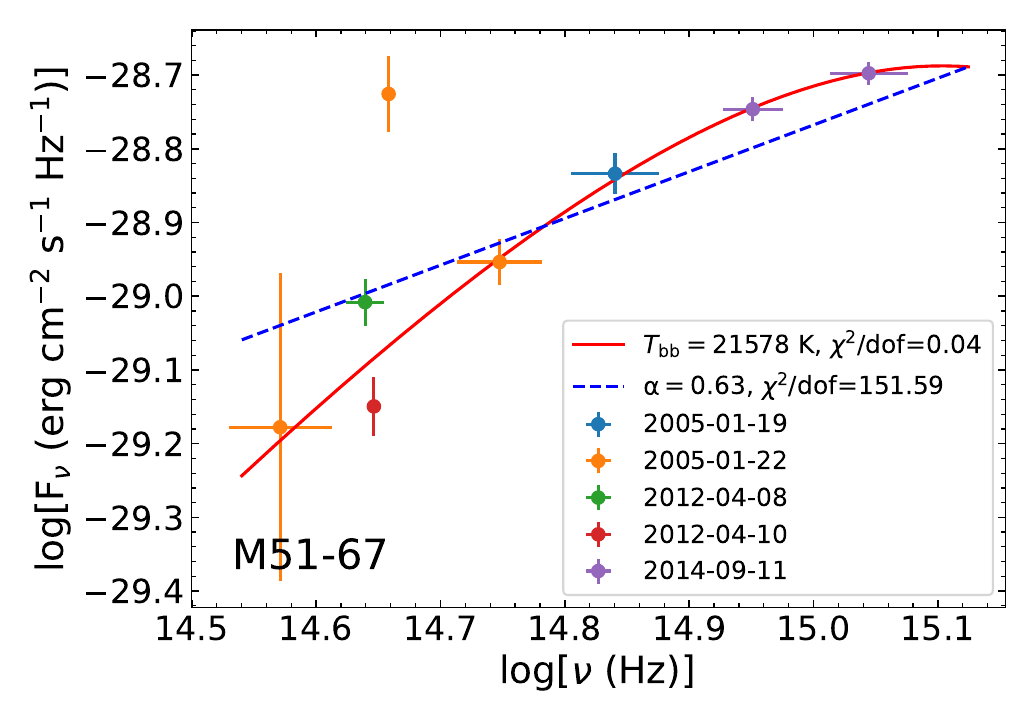}
\includegraphics[width=0.33\linewidth]{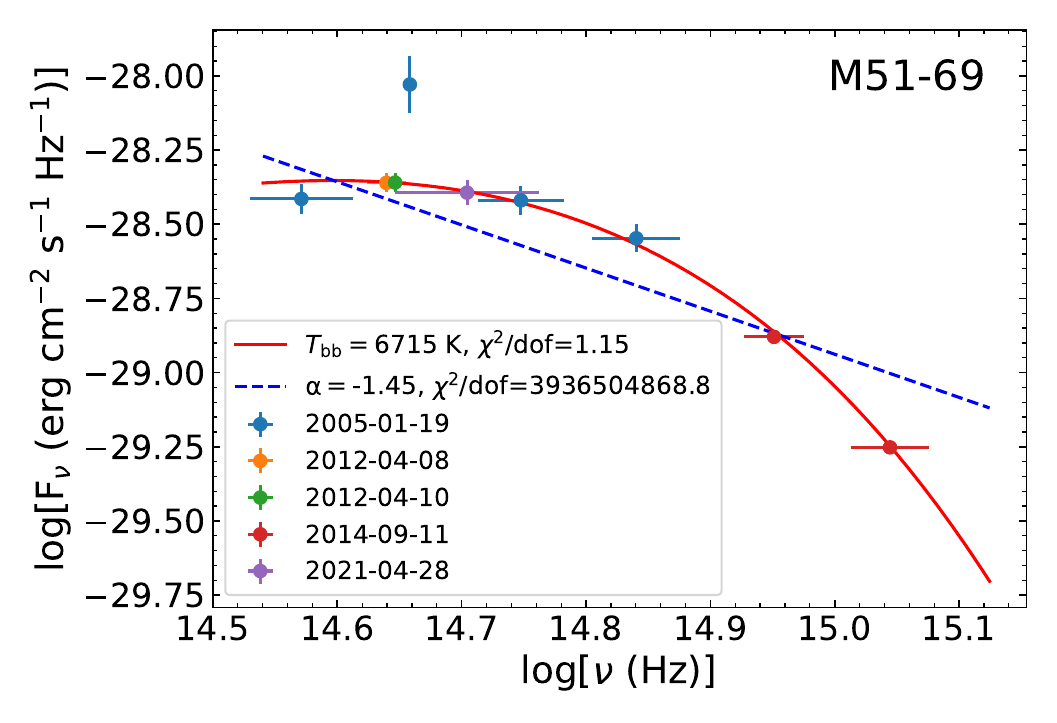}
\includegraphics[width=0.33\linewidth]{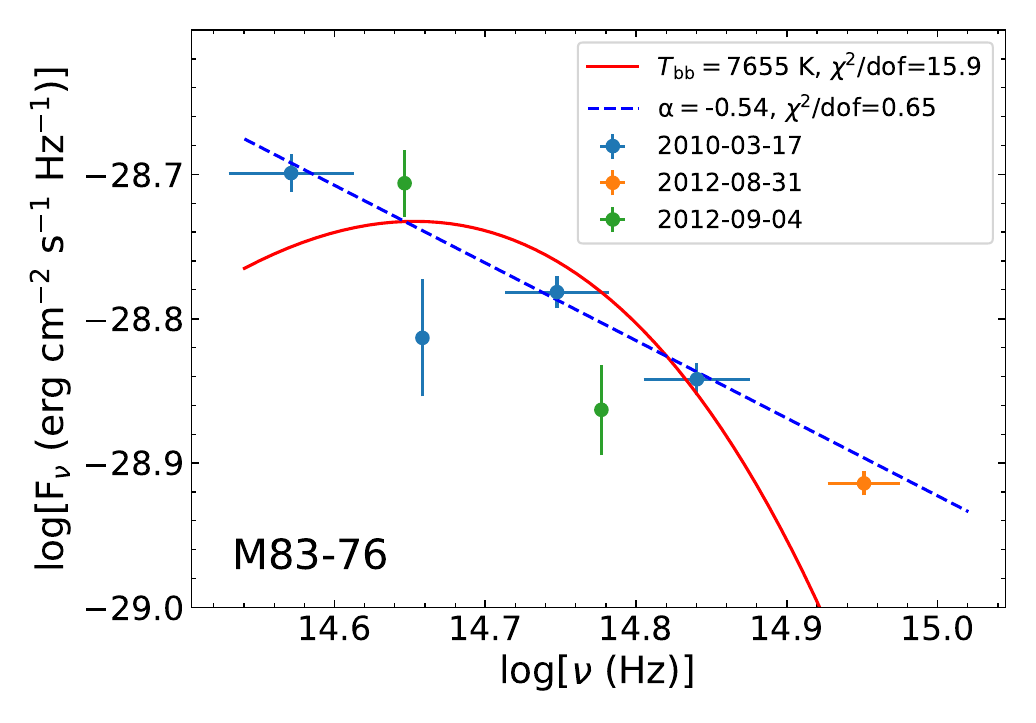}
\includegraphics[width=0.33\linewidth]{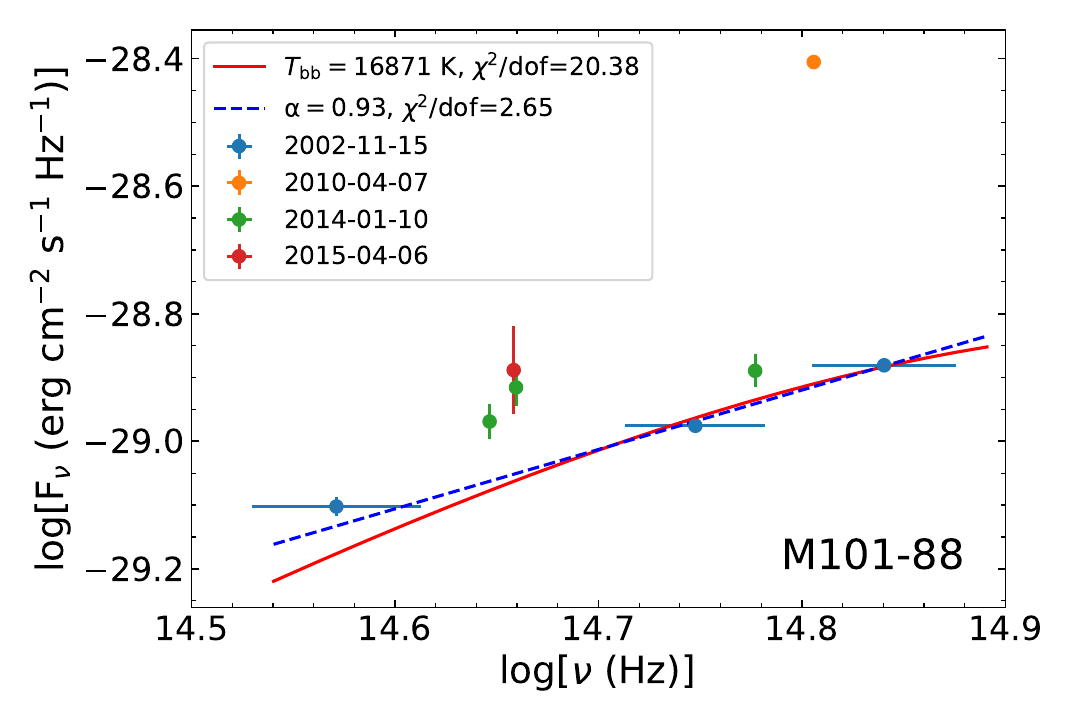}
\includegraphics[width=0.33\linewidth]{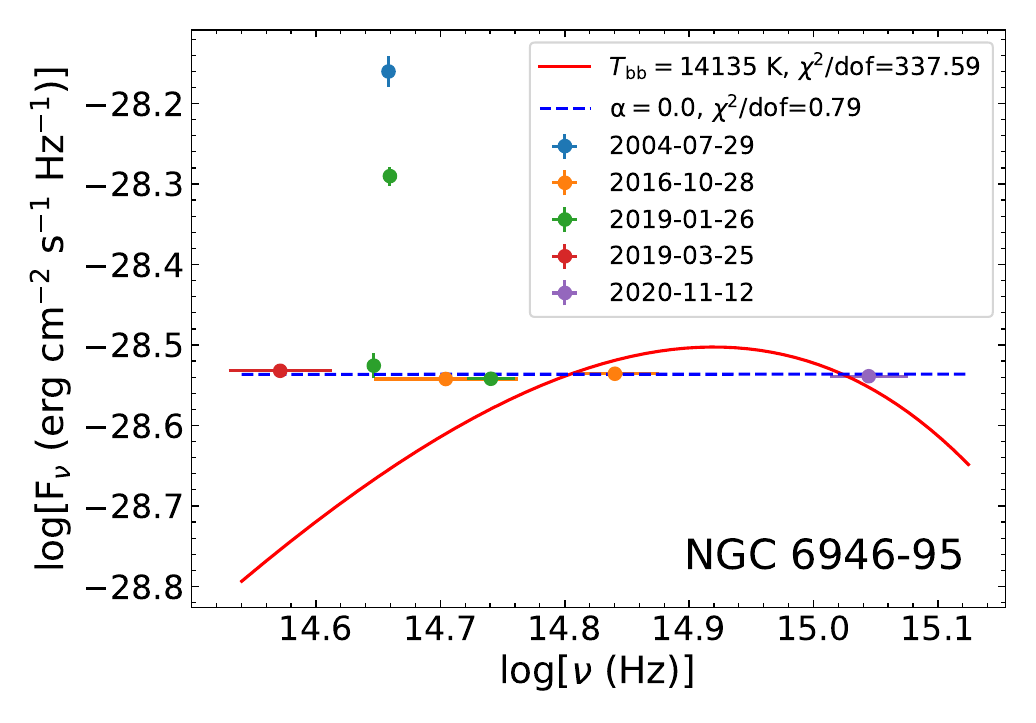}
\caption{SEDs of the candidate counterparts with model curves (power-law and blackbody, respectively). The best-fit power-law index $\alpha$ and blackbody temperature $T_{\rm bb}$ are shown in each plot. Narrow band fluxes are shown but not used for model fits.}
\label{fig:sed}
\end{figure*}

\subsection{Photometry and SED}

The HST photometry is performed using the Python package {\tt photutils}. 
In general, we chose a source aperture with a radius of 0\farcs15 and an annulus background aperture from 0\farcs4--0\farcs6\ around the source; 
for source NGC~6946-95, a radius of 0\farcs1\ is used to avoid a nearby source;
for M31-4, an aperture with a radius of 0\farcs1\ centered at the supposed counterpart is used.
Given the net count rate, the source flux or magnitude is then calculated using the Python version of {\tt synphot}, assuming a power-law spectrum for each object. 
Given multiple HST images in different broad bands, the power-law index is obtained iteratively until the output is consistent with the assumption. 
For the dereddened flux or absolute magnitude, only the Galactic extinction is considered \citep{Schlafly2011}. 
To compare with previous studies, we also inferred the extinction-corrected Vega magnitudes in the standard Johnson $UBV$ bands, converted from the closest HST bands (e.g., F336W to $U$, F435W to $B$, and F555W to $V$). 
The $U-B$ and $B-V$ colors and the absolute $V$-band magnitude are calculated and listed in Table~\ref{tab:johnson}.
The full photometry results of the candidate optical counterparts in each filter and each epoch are listed in the Appendix Table~\ref{tab:photometry}.

We fit the multi-band spectral energy density (SED in $F_{\nu}$) of each optical counterpart with a blackbody and power-law model, respectively.
The purpose of the SED fitting is not for accurate physical modeling, but to get a rough idea about the SED shape. 
For each source, we tried to select quasi-simultaneous observations (performed in the same day) to minimize the uncertainty of time variability.
The choice of observations does not affect the determination of the overall spectral shape, e.g., between a blackbody and a power-law, and has little impact on the estimated power-law index. 
The Galactic extinction along the line of sight is taken into account.
The SEDs with best-fit models are plotted in Figure~\ref{fig:sed}.
Only the broad band fluxes are used for the fit, but the narrow band fluxes are also plotted to examine if emission lines exist. 
The SEDs of six sources can be adequately fitted with a power-law model, as one can see with the $\chi^2$ and degree of freedom (dof) in the plot. 
For M31-4, M33-13, M51-67 and M51-69, the blackbody model is preferred.
We note that, these blackbody spectra may appear to be power-law-like with an additional extinction of $E(B-V) = 0.6 - 0.8$, which is higher than most of the extragalactic extinction inferred from nebular Balmer decrement \citep[see][where only 1 case out of 7 is possible]{Tao2011}.

As there is a chance of 10\% that the optical counterpart is outside the corrected error circle, we further examined the SEDs of optical sources outside it but within the original error circle. 
We found seven such sources with $m \lesssim 24$ around M31-3, M51-70, M51-71, M83-72, M83-73 (two), and M101-90. 
As a contrast, all of them show a blackbody-like SED (see Figure~\ref{fig:sed_cat_b} in Appendix~\ref{sec:append_sed}), justifying that the identification, from a statistical point of view, is valid.

\begin{deluxetable}{ccccrrr}
\label{tab:johnson}
\tablecaption{Optical magnitudes and colors in the standard Johnson bands of the optical counterparts.}
\tablehead{
\colhead{Source}&\colhead{$U$}&\colhead{$B$}&\colhead{$V$}&\colhead{$M_V$}&\colhead{$U-B$}&\colhead{$B-V$}}
\startdata
\noalign{\smallskip}  
\multicolumn{7}{c}{identified counterparts} \\
\noalign{\smallskip}\hline\noalign{\smallskip}  
M31-4&23.29&21.91&20.98&$-$3.48&1.38&0.93\\
M33-13&20.53&21.22&21.14&$-$3.43&$-$0.69&0.08\\
NGC\ 2403-35&21.61&22.67&22.85&$-$4.66&$-$1.06&$-$0.18\\
M81-40&21.21&21.86&21.71&$-$6.07&$-$0.65&0.15\\
M81-42&21.41&22.10&21.98&$-$5.80&$-$0.69&0.12\\
M51-67&22.57&23.64&23.76&$-$5.76&$-$1.07&$-$0.12\\
M51-69&22.65&22.85&22.43&$-$7.09&$-$0.20&0.42\\
M83-76&22.77&23.62&23.45&$-$4.86&$-$0.85&0.17\\
M101-88&23.18&23.62&23.94&$-$5.09&$-$0.44&$-$0.32\\
NGC\ 6946-95&21.81&22.89&22.76&$-$6.09&$-$1.08&0.13\\
\noalign{\smallskip}\hline\noalign{\smallskip}  
\multicolumn{7}{c}{$2\sigma$ upper limits for unidentified counterparts} \\
\noalign{\smallskip}\hline\noalign{\smallskip}  
M31-2&\nodata&\nodata&23.26&$-$1.20&\nodata&\nodata\\
M31-3&\nodata&\nodata&21.92&$-$2.54&\nodata&\nodata\\
M81-41&\nodata&\nodata&26.53&$-$1.25&\nodata&\nodata\\
M51-70&\nodata&\nodata&23.27&$-$6.25&\nodata&\nodata\\
M51-71&\nodata&\nodata&21.41&$-$8.11&\nodata&\nodata\\
M83-72&\nodata&\nodata&27.02&$-$1.29&\nodata&\nodata\\
M83-73&\nodata&\nodata&25.24&$-$3.07&\nodata&\nodata\\
M83-78&\nodata&\nodata&27.01&$-$1.21&\nodata&\nodata\\
M101-84&\nodata&\nodata&24.10&$-$4.93&\nodata&\nodata\\
M101-87&\nodata&\nodata&26.70&$-$2.33&\nodata&\nodata\\
M101-89&\nodata&\nodata&27.16&$-$1.87&\nodata&\nodata\\
M101-90&\nodata&\nodata&20.85&$-$8.18&\nodata&\nodata\\
\enddata
\end{deluxetable}

\section{Discussion} 
\label{sec:discuss}

Based on the sample of \citet{Zhou2019}, we identified a potential optical counterpart for 10 X-ray sources. 
For the rest 12 sources, no optical counterpart can be identified in the image.

Among the sample, two sources have been studied in optical, M51-67 \citep[M51 ULX-2;][]{Terashima2006} and M101-88 \citep[M101 ULX-1;][]{Kong2004, M101ref, Liu2009, Liu2013}. 
We obtained consistent results as those reported in the literature. 

\subsection{Comparison with ULXs}

\begin{figure}[tb]
\centering
\includegraphics[width=0.8\columnwidth]{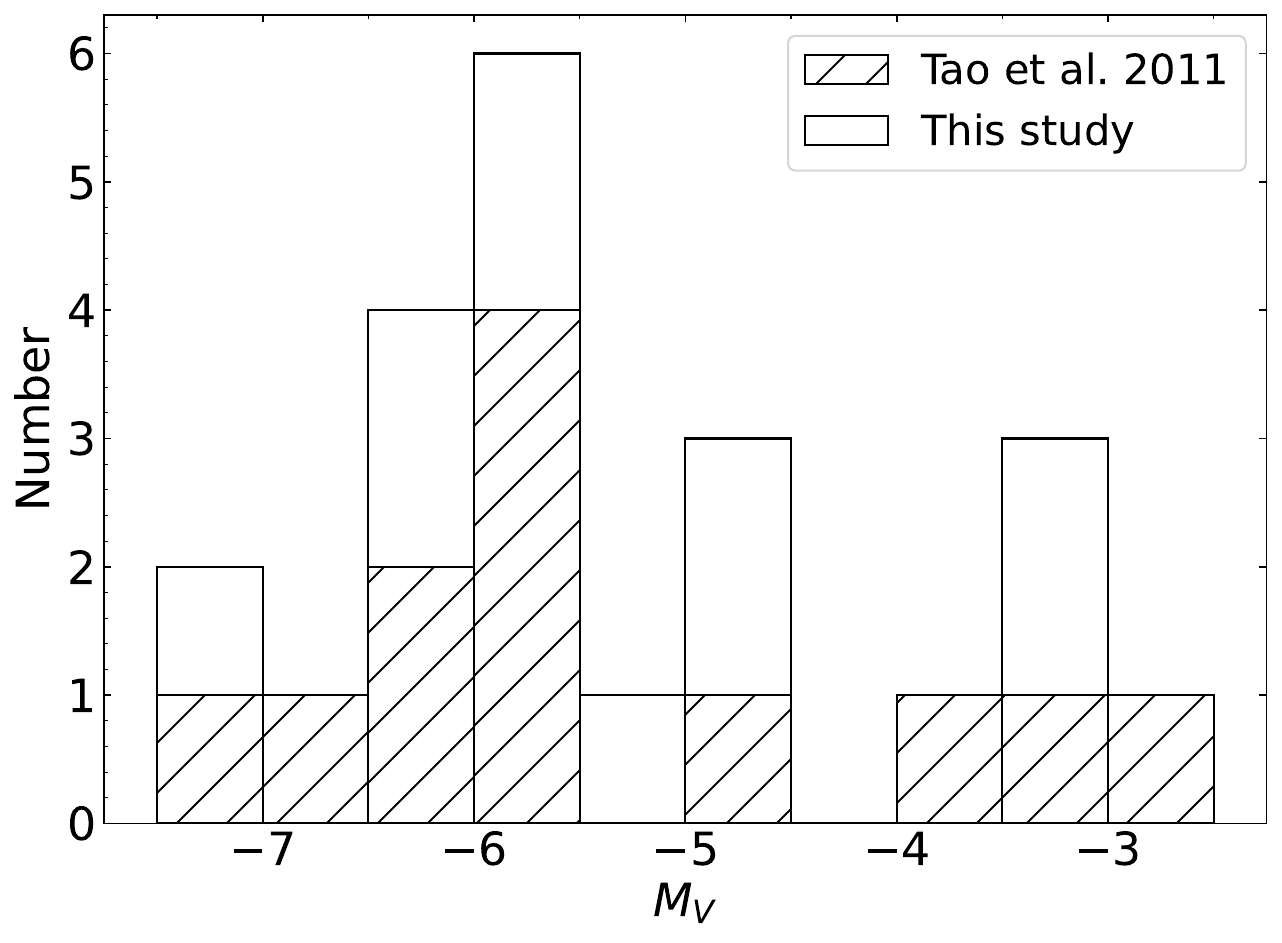}
\caption{Distributions of absolute visual magnitude of ULXs \citep{Tao2011,Gladstone2013} and the 9 objects in our sample with a unique identification of an optical counterpart. }
\label{fig:mv}
\end{figure}

The main purpose of this work is to test whether or not these objects are consistent with supercritical accretion systems viewed nearly edge-on. 
If the scenario is correct, we expect that they exhibit similar behaviors in optical as ULXs do, because the optical emission originating from the outer disk or companion star is nearly isotropic.
The optical counterparts of ULXs display an absolute visual magnitude in the range from $-3$ to $-7$ \citep{Tao2011, Gladstone2013}, and the 10 objects in our sample identified with an optical counterpart show a similar $M_V$ distribution (Figure~\ref{fig:mv}).
However, for the rest 12 objects without an obvious optical counterpart, they may be much fainter than ULXs in the optical band.
Their absolute $V$ magnitudes are estimated to have a 2$\sigma$ upper limit $\gtrsim -3$ for 8 objects.
This may suggest that the sample of \citet{Zhou2019} have a diverse population. 
In the following, we focus on the sub-sample with an optical identification. 

\begin{figure}
\centering
\includegraphics[width=0.8\columnwidth]{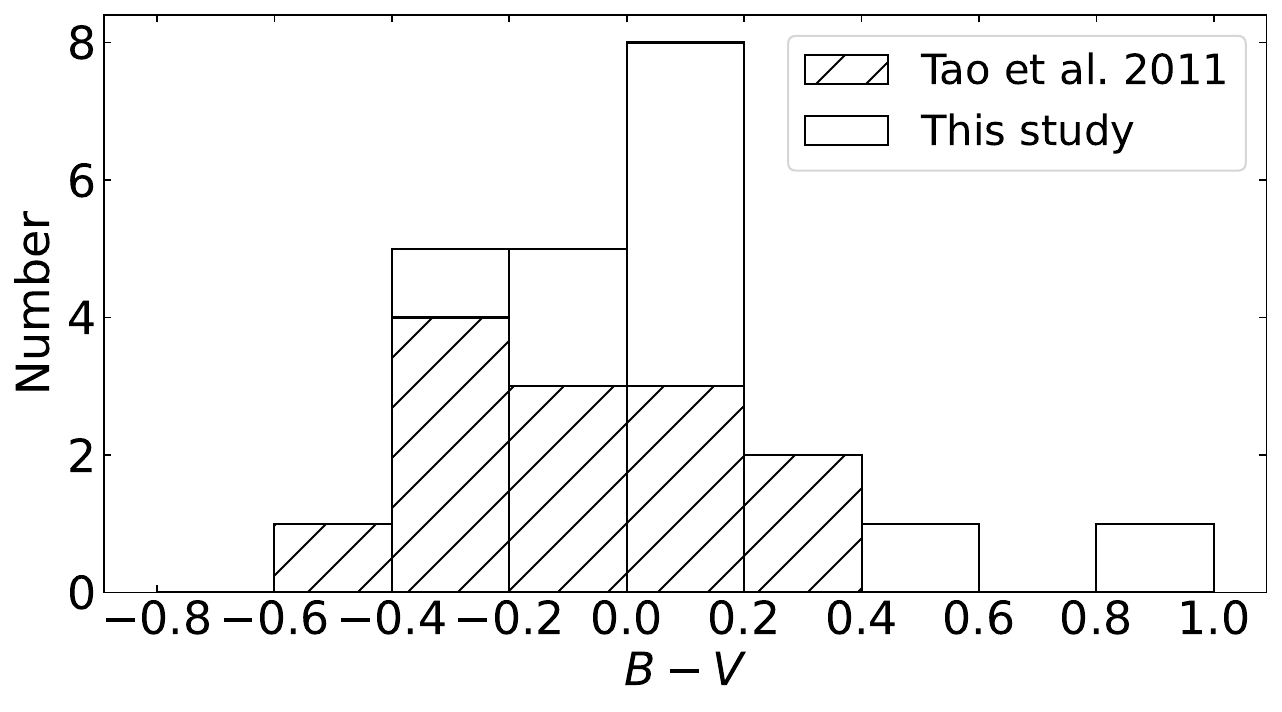}
\includegraphics[width=0.8\columnwidth]{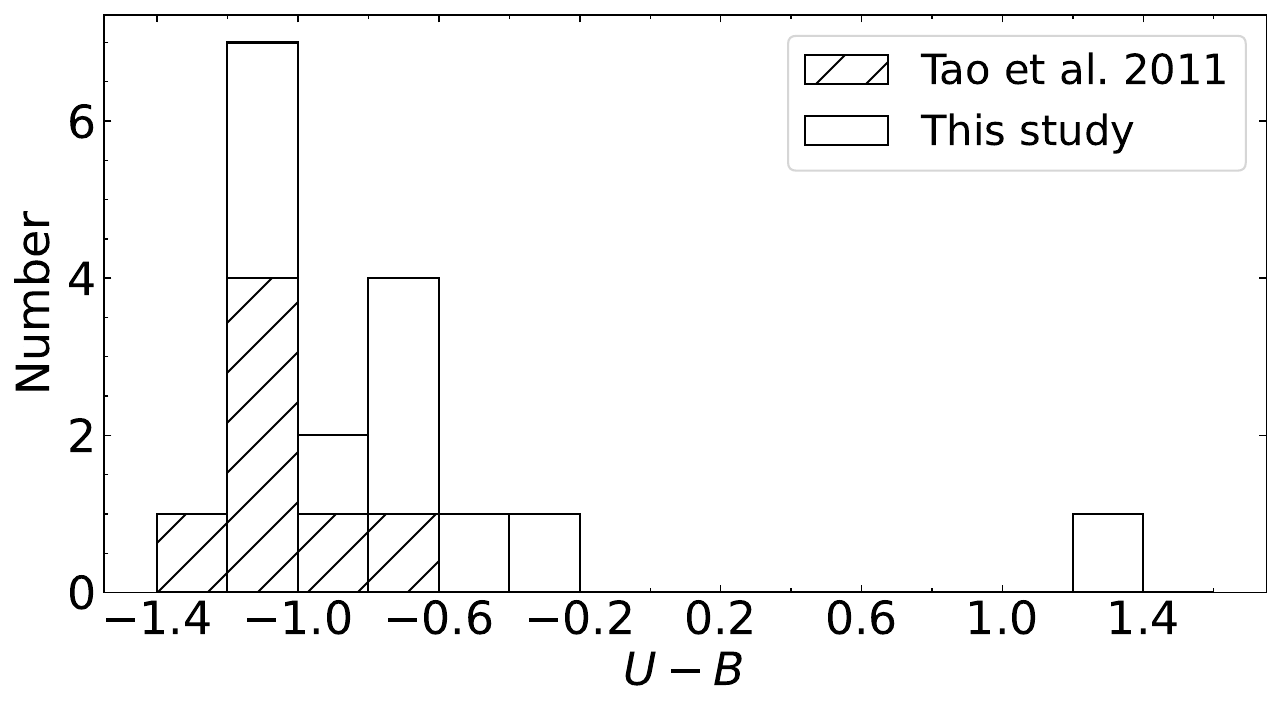}
\caption{Color index distribution for objects in this study compared with those in \citet{Tao2011}.}
\label{fig:color}
\end{figure}

The $B-V$ and $U-B$ colors are also compared between the two samples, see Figure~\ref{fig:color}. 
As one can see, both the $B-V$ and $U-B$ colors shows consistent distributions between the two samples, in spite of the small sample size.
The $U-B$ color is more sensitive to the correction of extinction, while the dust extinction in the host galaxies is not considered in this study.
For some ULXs (7 out of 13) associated with a nebula in \citet{Tao2011}, the extinction inferred from nebular Balmer decrement has a median excess of $E(B-V) = 0.13$ or $E(U-B) = 0.09$.
If the same amount of excessive extinction applies to all objects in our sample, the distributions between the two populations are also consistent.
M31-4 shows rather red colors deviating from other sources. 
Considering that this object may lie in a diffuse \halpha\ component, it is likely that there is excessive extinction not corrected.
Nevertheless, one should keep in mind the caveat of the small sample and limited statistics. 

\begin{figure}[tb]
\centering
\includegraphics[width=0.8\columnwidth]{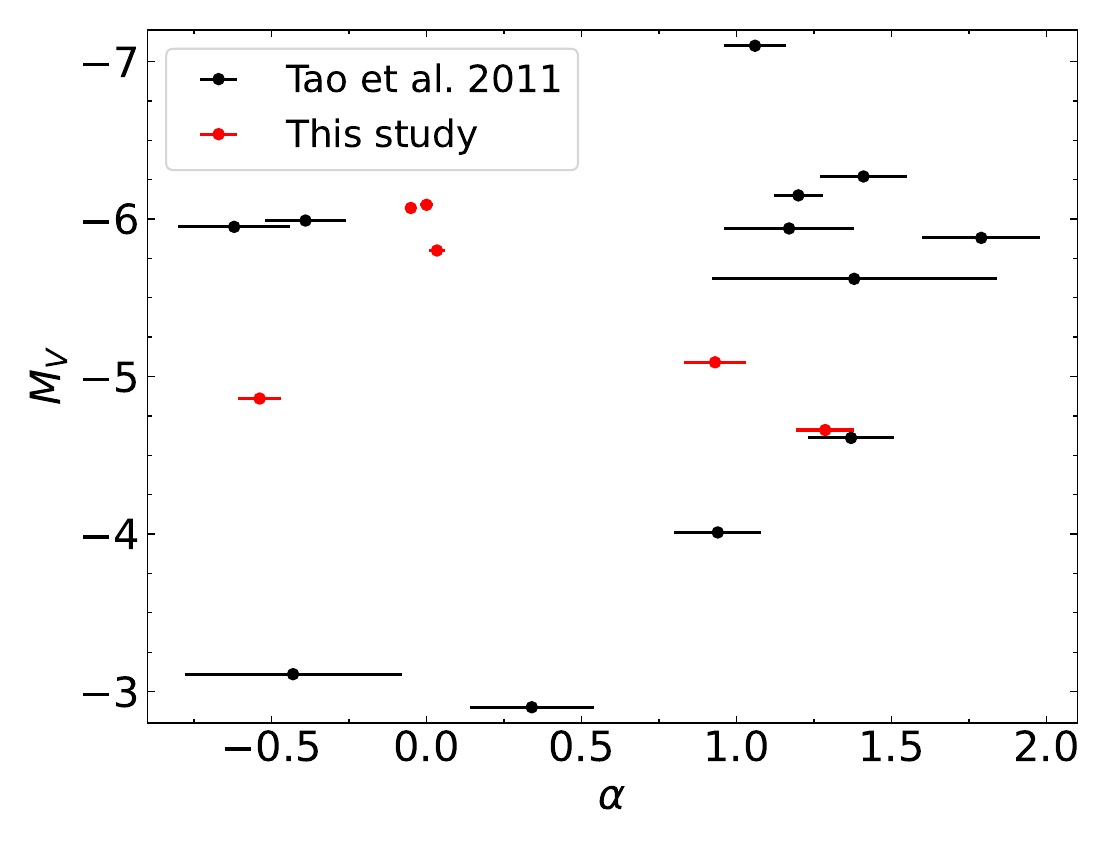}
\caption{$M_V$ vs.\ $\alpha$ for objects in this study compared with those in \citet{Tao2011}. For the same source with repeated observations in \citet{Tao2011}, the average is used.}
\label{fig:alpha}
\end{figure}

If the optical emission is driven by the accretion process, e.g., from the outer accretion disk due to X-ray irradiation, one expects temporal variability. 
We examined the variability in sources with repeated HST observations in the same filter.
For M101-88, multiple observations are available in F555W and F814W, showing strong time variability (see Table~\ref{tab:photometry} in Appendix) consistent with that reported in the literature \citep{Liu2009}, who argued that the optical variation is unlikely due to the companion based on its correlation with X-ray flux and the possible spectral type.
Both the magnitudes and color are found to be time variable, and the variation timescale is as short as 10~days, in favor of the optical emission being dominated by the accretion processes. 


\begin{figure}[tb]
\centering
\includegraphics[width=0.9\columnwidth]{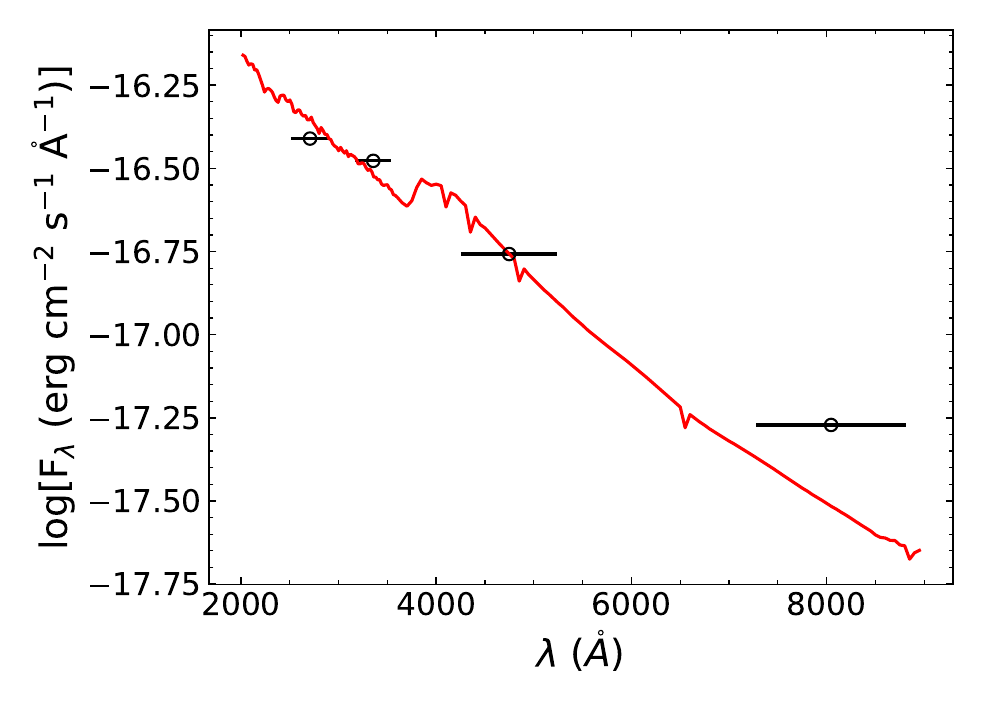}
\caption{SED of M33-13 compared with the template spectrum of a B5V main sequence star.}
\label{fig:13_B5V}
\end{figure}

Despite the flux variability, for sources with a power-law SED, we may make a direct comparison of the power-law index with the ULX sample. 
In Figure~\ref{fig:alpha}, we plot the $M_V - \alpha$ ($F_\nu \propto \nu^\alpha$) diagram and have a comparison with the ULX sample in \citet{Tao2011}, using the 6 objects that have a power-law like SED. 
Our sample is not contradicting the ULX population in the $\alpha \lesssim 1$ regime.
It seems like there is a missing population in our sample in the parameter space with $\alpha \gtrsim 1$ and $M_V \lesssim -5.5$, where most ULXs are located.
We note that $\alpha = 1/3$ if the optical emission originates from a standard multicolor blackbody spectrum \citep{1973A&A....24..337S}, and approaches 2 in the case of Rayleigh-Jeans approximation. 
No sources in our sample have such indices.
Apparently, there is no source from samples of this study in the latter case. 
For X-ray irradiation on the outer disk, the spectral index may vary between $-1$ and $2$ \citep{2009MNRAS.392.1106G}.
Our results are not in conflict with with the disk irradiation scenario, but no constraining conclusions can be drawn due to the small sample.

\subsection{Be White Dwarfs?}

It was argued in \citet{Zhou2019} that these sources were too luminous and too hot to be traditional supersoft X-ray sources, which are powered by nuclear burning on the surface of white dwarfs. 
However, there is a rare population of Be X-ray binaries containing a white dwarf \citep{Kahabka2006}, which may appear as luminous as $10^{38}$~\ergs\ in the form of a dominant thermal component with a temperature of $\sim$0.1~keV \citep{Marino2024}. 
We thus compared the colors and magnitudes of the objects in our sample to those of the late A to early O types, in a luminosity range from the main sequence to giant \citep{2000asqu.book.....C}.
M33-13 is the only object that shows a possible match.
We then further compared its SED with the stellar spectral templates in the Bruzual Atlas in Synphot, and found that its spectrum resembles a B5V star except in the F814W band, which shows an excess and can be explained as due to unresolved nearby structures (Figure~\ref{fig:13_B5V}). 
However, the absence of narrow band observations for M33-13 hinders a definitive determination of its nature.
We also note that M33-13 has an X-ray luminosity of only $4 \times 10^{36}$~\ergs\ (as explained in Section~\ref{sec:sample}).
More X-ray observations are needed to further determine its nature. 

Three other sources (M31-4, M51-67 and M51-69) also show a blackbody-like SED.
The color and magnitude are reminiscent of a B-type Ib supergiant for M51-67, consistent with the identification in \citet{Terashima2006}, an F-type Ia supergiant for M51-69, and an G-type Ib supergiant for M31-4.
Most of them appear like point-like sources in the HST image, indicating that their spatial extent is less than 4~pc ($< 0\farcs1$) at a distance of 8~Mpc. 
These suggest, along with the color and magnitude, that they are unlikely young star clusters.

\subsection{Emission lines}

Many optical counterparts display excessive line emission compared with the continuum, as can be seen from the SEDs in Figure~\ref{fig:sed} that some narrow-band fluxes are significantly above the model.
These are \halpha\ emission lines seen in sources NGC~2403-35, M81-40, M51-67, M51-69, M101-88 and NGC~6946-95, and \HeII\ and \SII\ emission lines also seen in M101-88;
the excess of line emission in M51-67 and M101-88 is consistent with previous spectroscopic results \citep{Kuntz2005, Terashima2006, Abolmasov2007, Liu2013}. 
The nature of the line emission is unclear,
but is also seen in soft ULXs like NGC 55 ULX \citep{Zhou_2023_ApJ_955_61}. 
Future spectroscopic observations are needed to investigate whether the \halpha\ profile is narrow or broad and if the line flux is variable. 

\section{Summary}

To summarize, the optical study of this work revealed a diverse population. 
The optical emission of these X-ray sources exhibit a wide range, with $M_V$ from $>-3$ to about $-7$ with some upper limits of $M_V > -1.2$. 
Many sources show a power-law optical SED while a few of them show a blackbody-body SED, with one consistent with a B type main sequence and the other three consistent with supergiants.
This suggests that at least one of them is a candidate Be white dwarf system. 
The magnitudes, colors, and power-law spectral indices of some sources are consistent with those seen ULXs. 
Strong time variability on timescales as short as 10 days indicates that they are powered by accretion. 
Thus, these objects are interesting candidates for supercritical accretion systems and deserve future spectroscopic investigations. 

\begin{acknowledgments}
We thank the anonymous referee for useful comments. HF acknowledges funding support from the National Natural Science Foundation of China under the grant 12025301.
This paper employs a list of Chandra datasets, obtained by the Chandra X-ray Observatory, contained in \dataset[DOI: 10.25574/cdc.403]{https://doi.org/10.25574/cdc.403}, and a list of HST data obtained from the Mikulski Archive for Space Telescopes (MAST) at the Space Telescope Science Institute, contained in \dataset[DOI: 10.17909/0sv9-me14]{http://dx.doi.org/10.17909/0sv9-me14}. 
\end{acknowledgments}

\vspace{5mm}
\facilities{HST, CXO}

\bibliography{refs}
\bibliographystyle{aasjournal}


\appendix
\section{HST observations and photometry results}
\label{sec:append}

\startlongtable
\tabletypesize\tiny
\begin{deluxetable}{llllll}
\tablecaption{HST Observations and photometry of the optical counterparts.}
\label{tab:photometry}
\tablehead{
\colhead{ID} & \colhead{Date} & \colhead{ObsID} & \colhead{Instrument} & \colhead{Exposure (s)} & \colhead{ST mag}
}
\startdata
M31-4
&2006-02-10&J9JU01010&ACS/WFC1/F435W&4360.0&21.331$\pm$0.062\\
&2007-01-10&J9JU06010&ACS/WFC1/F435W&4672.0&21.324$\pm$0.063\\
&2010-01-21&JB9D15010&ACS/WFC1/F435W&4360.0&21.282$\pm$0.059\\
&2010-07-20&JB9D20010&ACS/WFC1/F435W&4360.0&21.316$\pm$0.060\\
&2010-07-21&IBF310030&WFC3/UVIS2/F336W&1250.0&22.650$\pm$0.137\\
&2010-07-21&IBF310040&WFC3/UVIS2/F275W&925.0&$>$23.222\\
&2010-07-24&JBF312010&ACS/WFC1/F814W&1520.0&21.282$\pm$0.078\\
&2010-07-24&JBF312020&ACS/WFC1/F475W&1720.0&21.159$\pm$0.054\\
&2010-12-21&IBIR01020&WFC3/UVIS2/F373N&2700.0&23.002$\pm$0.502\\
&2010-12-21&IBIR02020&WFC3/UVIS2/F656N&2700.0&21.085$\pm$0.084\\
&2010-12-21&IBIR03020&WFC3/UVIS2/F658N&2700.0&20.997$\pm$0.070\\
&2010-12-22&JBF307010&ACS/WFC1/F814W&1715.0&21.294$\pm$0.093\\
&2010-12-22&JBF307020&ACS/WFC1/F475W&1900.0&21.176$\pm$0.053\\
&2010-12-22&IBIR04020&WFC3/UVIS2/F502N&2700.0&21.117$\pm$0.066\\
&2010-12-23&IBIR05020&WFC3/UVIS2/F502N&2700.0&21.082$\pm$0.065\\
&2010-12-25&IBIR06020&WFC3/UVIS2/F390M&2700.0&23.150$\pm$0.227\\
&2010-12-26&IBIR07020&WFC3/UVIS2/F547M&2700.0&20.993$\pm$0.056\\
&2011-01-02&IBIR08020&WFC3/UVIS2/F665N&2700.0&21.085$\pm$0.057\\
&2015-08-26&ICJ204050&WFC3/UVIS2/F225W&2340.0&$>$23.028\\
&2015-08-26&ICJ204060&WFC3/UVIS2/F225W&2472.0&$>$23.015\\
&2015-08-26&ICJ204070&WFC3/UVIS2/F225W&2469.0&$>$23.022\\
&2015-08-26&ICJ204080&WFC3/UVIS2/F336W&2276.0&23.184$\pm$0.195\\
\hline
M33-13
&2017-08-13&IDB642030&WFC3/UVIS2/F336W&1250.0&20.247$\pm$0.007\\
&2017-08-13&IDB642040&WFC3/UVIS2/F275W&890.0&20.026$\pm$0.013\\
&2017-08-13&IDB641030&WFC3/UVIS2/F336W&1250.0&20.093$\pm$0.007\\
&2017-08-13&IDB641040&WFC3/UVIS2/F275W&890.0&19.926$\pm$0.012\\
&2018-01-20&JDB638010&ACS/WFC1/F814W&1503.0&22.072$\pm$0.008\\
&2018-01-21&JDB638020&ACS/WFC1/F475W&1705.0&20.786$\pm$0.006\\
&2018-01-21&JDB639010&ACS/WFC1/F814W&1503.0&22.086$\pm$0.007\\
&2018-01-21&JDB639020&ACS/WFC1/F475W&1705.0&20.803$\pm$0.006\\
\hline
NGC\ 2403-35
&2004-08-17&J90ZX1010&ACS/WFC1/F475W&1200.0&22.326$\pm$0.017\\
&2004-08-17&J90ZX1020&ACS/WFC1/F606W&700.0&23.115$\pm$0.040\\
&2004-08-17&J90ZX1030&ACS/WFC1/F814W&700.0&24.255$\pm$0.072\\
&2004-08-17&J90ZX1040&ACS/WFC1/F658N&1300.0&22.539$\pm$0.068\\
&2005-03-29&J96R27020&ACS/WFC1/F658N&1400.0&22.519$\pm$0.056\\
&2005-03-29&J96R27010&ACS/WFC1/F814W&750.0&24.224$\pm$0.073\\
&2005-10-17&J9H803010&ACS/WFC1/F435W&1248.0&22.035$\pm$0.017\\
&2005-10-17&J9H803020&ACS/WFC1/F606W&1248.0&23.018$\pm$0.041\\
&2019-08-20&JDXK26010&ACS/WFC1/F814W&2456.0&24.187$\pm$0.070\\
\hline
M81-40
&2003-09-18&J8MX18010&ACS/WFC1/F658N&700.0&20.762$\pm$0.020\\
&2004-09-15&J90LA7010&ACS/WFC1/F814W&1650.0&22.543$\pm$0.032\\
&2006-03-21&J9EL15010&ACS/WFC1/F435W&1200.0&21.237$\pm$0.008\\
&2006-03-21&J9EL15020&ACS/WFC1/F606W&1200.0&21.904$\pm$0.014\\
\hline
M81-42
&2004-09-15&J90LA9010&ACS/WFC1/F814W&1650.0&22.862$\pm$0.009\\
&2006-03-22&J9EL18010&ACS/WFC1/F435W&1200.0&21.477$\pm$0.008\\
&2006-03-22&J9EL18020&ACS/WFC1/F606W&1200.0&22.165$\pm$0.006\\
&2010-01-08&JB6K10010&ACS/WFC1/F555W&2000.0&21.936$\pm$0.007\\
\hline
M51-67
&2005-01-19&J97C43N2Q&ACS/WFC1/F435W&680.0&22.977$\pm$0.070\\
&2005-01-19&J97C43N3Q&ACS/WFC1/F555W&340.0&23.620$\pm$0.067\\
&2005-01-19&J97C43N5Q&ACS/WFC1/F814W&340.0&25.213$\pm$0.159\\
&2005-01-19&J97C43N7Q&ACS/WFC1/F658N&680.0&23.750$\pm$0.165\\
&2005-01-19&J97C44NAQ&ACS/WFC1/F435W&680.0&23.002$\pm$0.090\\
&2005-01-19&J97C44NBQ&ACS/WFC1/F555W&340.0&23.618$\pm$0.420\\
&2005-01-19&J97C44NDQ&ACS/WFC1/F814W&340.0&25.239$\pm$0.145\\
&2005-01-19&J97C44NFQ&ACS/WFC1/F658N&680.0&23.866$\pm$0.160\\
&2005-01-22&J97C41XJQ&ACS/WFC1/F435W&680.0&22.872$\pm$0.064\\
&2005-01-22&J97C41XKQ&ACS/WFC1/F555W&340.0&23.684$\pm$0.074\\
&2005-01-22&J97C41XMQ&ACS/WFC1/F814W&340.0&25.269$\pm$0.156\\
&2005-01-22&J97C41XOQ&ACS/WFC1/F658N&680.0&23.494$\pm$0.242\\
&2005-01-22&J97C42XRQ&ACS/WFC1/F435W&680.0&22.774$\pm$0.053\\
&2005-01-22&J97C42XSQ&ACS/WFC1/F555W&340.0&23.741$\pm$0.077\\
&2005-01-22&J97C42XUQ&ACS/WFC1/F814W&340.0&25.194$\pm$0.522\\
&2005-01-22&J97C42XWQ&ACS/WFC1/F658N&680.0&23.614$\pm$0.128\\
&2012-04-08&IBVX06010&WFC3/UVIS2/F689M&1000.0&24.415$\pm$0.080\\
&2012-04-10&IBVX03010&WFC3/UVIS2/F673N&5400.0&24.733$\pm$0.100\\
&2014-09-11&ICD401010&WFC3/UVIS2/F275W&7147.0&21.611$\pm$0.038\\
&2014-09-11&ICD401020&WFC3/UVIS2/F336W&4360.0&22.203$\pm$0.042\\
\hline
M51-69
&2005-01-19&J97C33O8Q&ACS/WFC1/F435W&680.0&22.259$\pm$0.117\\
&2005-01-19&J97C33O9Q&ACS/WFC1/F555W&340.0&22.402$\pm$0.125\\
&2005-01-19&J97C33OBQ&ACS/WFC1/F814W&340.0&23.262$\pm$0.126\\
&2005-01-19&J97C33ODQ&ACS/WFC1/F658N&680.0&21.872$\pm$0.238\\
&2005-01-20&J97C31RRQ&ACS/WFC1/F435W&680.0&22.307$\pm$0.122\\
&2005-01-20&J97C31RSQ&ACS/WFC1/F555W&340.0&22.409$\pm$0.105\\
&2005-01-20&J97C31RUQ&ACS/WFC1/F814W&340.0&23.270$\pm$0.132\\
&2005-01-20&J97C31RWQ&ACS/WFC1/F658N&680.0&21.915$\pm$0.122\\
&2005-01-21&J97C32UTQ&ACS/WFC1/F435W&680.0&22.269$\pm$0.147\\
&2005-01-21&J97C32UUQ&ACS/WFC1/F555W&340.0&22.402$\pm$0.136\\
&2005-01-21&J97C32UWQ&ACS/WFC1/F814W&340.0&23.268$\pm$0.140\\
&2005-01-21&J97C32UYQ&ACS/WFC1/F658N&680.0&21.878$\pm$0.069\\
&2005-01-22&J97C34XBQ&ACS/WFC1/F435W&680.0&22.238$\pm$0.114\\
&2005-01-22&J97C34XCQ&ACS/WFC1/F555W&340.0&22.397$\pm$0.111\\
&2005-01-22&J97C34XEQ&ACS/WFC1/F814W&340.0&23.262$\pm$0.129\\
&2005-01-22&J97C34XGQ&ACS/WFC1/F658N&680.0&21.828$\pm$0.062\\
&2012-04-08&IBVXA6010&WFC3/UVIS2/F689M&1000.0&22.795$\pm$0.081\\
&2012-04-10&IBVX02010&WFC3/UVIS2/F673N&5400.0&22.758$\pm$0.081\\
&2012-04-12&IBVX04010&WFC3/UVIS2/F673N&5400.0&22.809$\pm$0.079\\
&2014-09-11&ICD401010&WFC3/UVIS2/F275W&7147.0&23.002$\pm$0.039\\
&2014-09-11&ICD401020&WFC3/UVIS2/F336W&4360.0&22.535$\pm$0.026\\
&2021-04-28&JEJK01010&ACS/WFC1/F606W&2208.0&22.552$\pm$0.106\\
&2021-04-29&JEJK02010&ACS/WFC1/F814W&2208.0&23.318$\pm$0.128\\
\hline
M83-76
&2004-08-07&J8PH0H030&ACS/WFC1/F814W&430.0&23.972$\pm$0.035\\
&2004-08-07&J8PH0H010&ACS/WFC1/F435W&680.0&22.922$\pm$0.031\\
&2004-08-07&J8PH0H020&ACS/WFC1/F555W&680.0&23.359$\pm$0.032\\
&2010-03-17&JB6WB1010&ACS/WFC1/F435W&1450.0&23.000$\pm$0.028\\
&2010-03-17&JB6WB1020&ACS/WFC1/F555W&1400.0&23.313$\pm$0.028\\
&2010-03-17&JB6WB1030&ACS/WFC1/F814W&1500.0&23.999$\pm$0.033\\
&2010-03-17&JB6WB1040&ACS/WFC1/F658N&1050.0&23.834$\pm$0.101\\
&2010-03-17&JB6WB3010&ACS/WFC1/F658N&1980.0&23.685$\pm$0.062\\
&2012-08-31&IBQC07030&WFC3/UVIS2/F814W&1379.0&23.974$\pm$0.033\\
&2012-08-31&IBQC07040&WFC3/UVIS2/F438W&1799.0&23.048$\pm$0.030\\
&2012-08-31&IBQC07050&WFC3/UVIS2/F657N&1799.0&23.869$\pm$0.083\\
&2012-08-31&IBQC07060&WFC3/UVIS2/F336W&2579.0&22.623$\pm$0.021\\
&2012-09-04&IBQC08040&WFC3/UVIS2/F502N&2982.0&23.365$\pm$0.078\\
&2012-09-04&IBQC08050&WFC3/UVIS2/F547M&2682.0&23.441$\pm$0.026\\
&2012-09-04&IBQC08060&WFC3/UVIS2/F673N&2262.0&23.625$\pm$0.058\\
\hline
M101-88
&2002-11-15&J8D601011&ACS/WFC1/F435W&900.0&23.093$\pm$0.018\\
&2002-11-15&J8D601021&ACS/WFC1/F555W&720.0&23.794$\pm$0.023\\
&2002-11-15&J8D601031&ACS/WFC1/F814W&720.0&25.004$\pm$0.037\\
&2006-12-23&J9O401010&ACS/WFC1/F814W&724.0&24.880$\pm$0.030\\
&2006-12-23&J9O401020&ACS/WFC1/F555W&1330.0&23.642$\pm$0.016\\
&2006-12-24&J9O402010&ACS/WFC1/F814W&724.0&24.858$\pm$0.031\\
&2006-12-24&J9O402020&ACS/WFC1/F555W&1330.0&23.662$\pm$0.017\\
&2006-12-25&J9O403010&ACS/WFC1/F814W&724.0&24.909$\pm$0.030\\
&2006-12-25&J9O403020&ACS/WFC1/F555W&1330.0&23.695$\pm$0.016\\
&2006-12-26&J9O404010&ACS/WFC1/F814W&724.0&24.948$\pm$0.035\\
&2006-12-26&J9O404020&ACS/WFC1/F555W&1330.0&23.687$\pm$0.016\\
&2006-12-28&J9O405010&ACS/WFC1/F814W&724.0&24.916$\pm$0.033\\
&2006-12-28&J9O405020&ACS/WFC1/F555W&1330.0&23.639$\pm$0.016\\
&2006-12-30&J9O406010&ACS/WFC1/F814W&724.0&24.898$\pm$0.034\\
&2006-12-30&J9O406020&ACS/WFC1/F555W&1330.0&23.713$\pm$0.017\\
&2007-01-01&J9O407010&ACS/WFC1/F814W&724.0&24.966$\pm$0.032\\
&2007-01-01&J9O407020&ACS/WFC1/F555W&1330.0&23.744$\pm$0.026\\
&2007-01-04&J9O408010&ACS/WFC1/F814W&724.0&24.960$\pm$0.034\\
&2007-01-04&J9O408020&ACS/WFC1/F555W&1330.0&23.738$\pm$0.016\\
&2007-01-07&J9O409010&ACS/WFC1/F814W&724.0&24.943$\pm$0.033\\
&2007-01-07&J9O409020&ACS/WFC1/F555W&1330.0&23.782$\pm$0.018\\
&2007-01-11&J9O410010&ACS/WFC1/F814W&724.0&24.826$\pm$0.031\\
&2007-01-11&J9O410020&ACS/WFC1/F555W&1330.0&23.679$\pm$0.016\\
&2007-01-17&J9O411010&ACS/WFC1/F814W&724.0&24.690$\pm$0.026\\
&2007-01-17&J9O411020&ACS/WFC1/F555W&1330.0&23.529$\pm$0.015\\
&2007-01-21&J9O412010&ACS/WFC1/F814W&724.0&24.777$\pm$0.028\\
&2007-01-21&J9O412020&ACS/WFC1/F555W&1330.0&23.594$\pm$0.015\\
&2010-04-07&IB3P10010&WFC3/UVIS2/F469N&6106.0&22.077$\pm$0.023\\
&2013-03-03&IC1371Q5Q&WFC3/UVIS2/F555W&472.0&23.614$\pm$0.063\\
&2013-03-03&IC1371Q7Q&WFC3/UVIS2/F814W&472.0&24.840$\pm$0.276\\
&2013-03-10&IC1372XSQ&WFC3/UVIS2/F555W&472.0&23.665$\pm$0.022\\
&2013-03-10&IC1372XUQ&WFC3/UVIS2/F814W&472.0&24.971$\pm$0.045\\
&2013-03-11&JC1373010&ACS/WFC1/F555W&750.0&23.803$\pm$0.026\\
&2013-03-18&JC1374010&ACS/WFC1/F555W&750.0&23.672$\pm$0.024\\
&2014-01-10&ICAE01010&WFC3/UVIS2/F657N&2150.0&24.084$\pm$0.072\\
&2014-01-10&ICAE01020&WFC3/UVIS2/F502N&2600.0&23.431$\pm$0.065\\
&2014-01-10&ICAE01030&WFC3/UVIS2/F673N&2600.0&24.282$\pm$0.067\\
&2014-01-10&ICAE01040&WFC3/UVIS2/F547M&710.0&23.927$\pm$0.046\\
&2015-04-06&JCOY02010&ACS/WFC1/F814W&2522.0&24.716$\pm$0.022\\
&2015-04-06&JCOY02020&ACS/WFC1/F555W&3124.0&23.589$\pm$0.011\\
&2015-04-06&JCOY02030&ACS/WFC1/F435W&4614.0&22.986$\pm$0.008\\
&2015-04-06&JCOY02050&ACS/WFC1/F658N&1397.0&24.021$\pm$0.174\\
&2020-04-23&IE3301010&WFC3/UVIS2/F300X&2560.0&26.828$\pm$0.515\\
&2020-06-10&IE3303010&WFC3/UVIS2/F300X&2560.0&22.463$\pm$0.015\\
\hline
NGC\ 6946-95
&2004-07-29&J8MXD8SXQ&ACS/WFC1/F814W&120.0&23.576$\pm$0.039\\
&2004-07-29&J8MXD8010&ACS/WFC1/F658N&700.0&22.201$\pm$0.047\\
&2016-10-28&JD9G01010&ACS/WFC1/F435W&5610.0&22.254$\pm$0.012\\
&2016-10-28&JD9G01020&ACS/WFC1/F606W&4478.0&22.978$\pm$0.013\\
&2019-01-26&IDK404030&WFC3/UVIS2/F657N&2826.0&22.520$\pm$0.030\\
&2019-01-26&IDK404040&WFC3/UVIS2/F673N&3853.0&23.174$\pm$0.039\\
&2019-01-26&IDK404050&WFC3/UVIS2/F547M&1473.0&22.751$\pm$0.022\\
&2019-01-26&IDK407030&WFC3/UVIS2/F657N&2826.0&22.462$\pm$0.029\\
&2019-01-26&IDK407040&WFC3/UVIS2/F673N&3853.0&23.392$\pm$0.044\\
&2019-01-27&IDK407050&WFC3/UVIS2/F547M&1473.0&22.796$\pm$0.023\\
&2019-03-25&JDXK36010&ACS/WFC1/F814W&1788.0&23.607$\pm$0.017\\
&2019-07-09&JDXK86010&ACS/WFC1/F814W&2284.671997&23.614$\pm$0.016\\
&2020-11-12&IE3Y07030&WFC3/UVIS2/F336W&2640.0&21.488$\pm$0.017\\
&2020-11-12&IE3Y07040&WFC3/UVIS2/F275W&5692.0&21.247$\pm$0.022\\
\enddata
\end{deluxetable}

\section{SEDs of spurious optical counterparts}
\label{sec:append_sed}


\begin{figure*}
\centering
\includegraphics[width=0.315\linewidth]{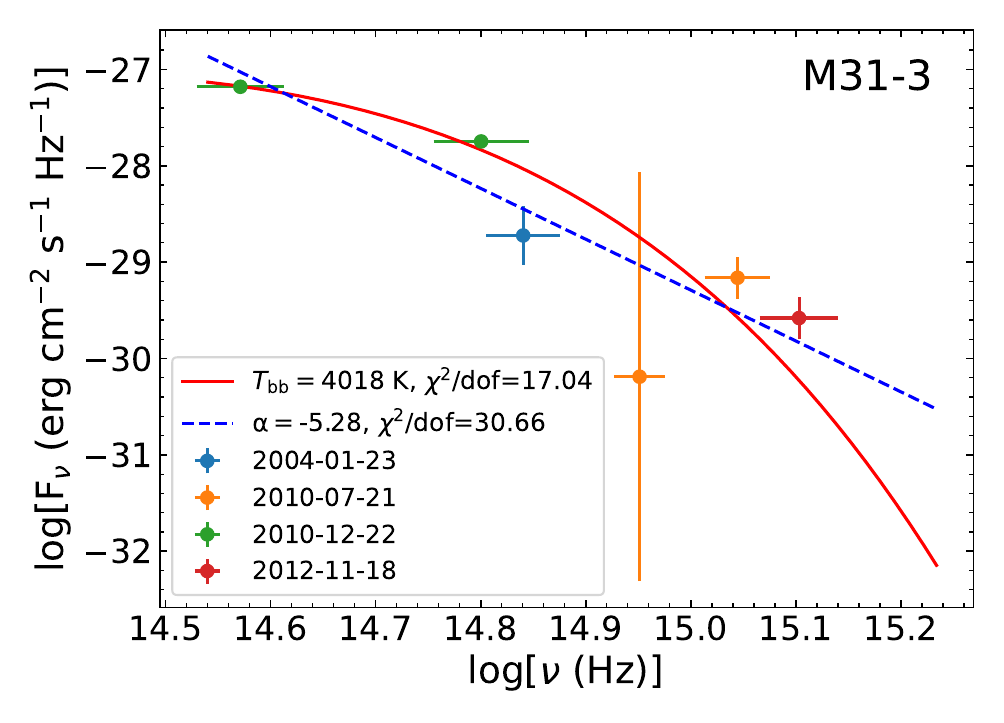}
\includegraphics[width=0.33\linewidth]{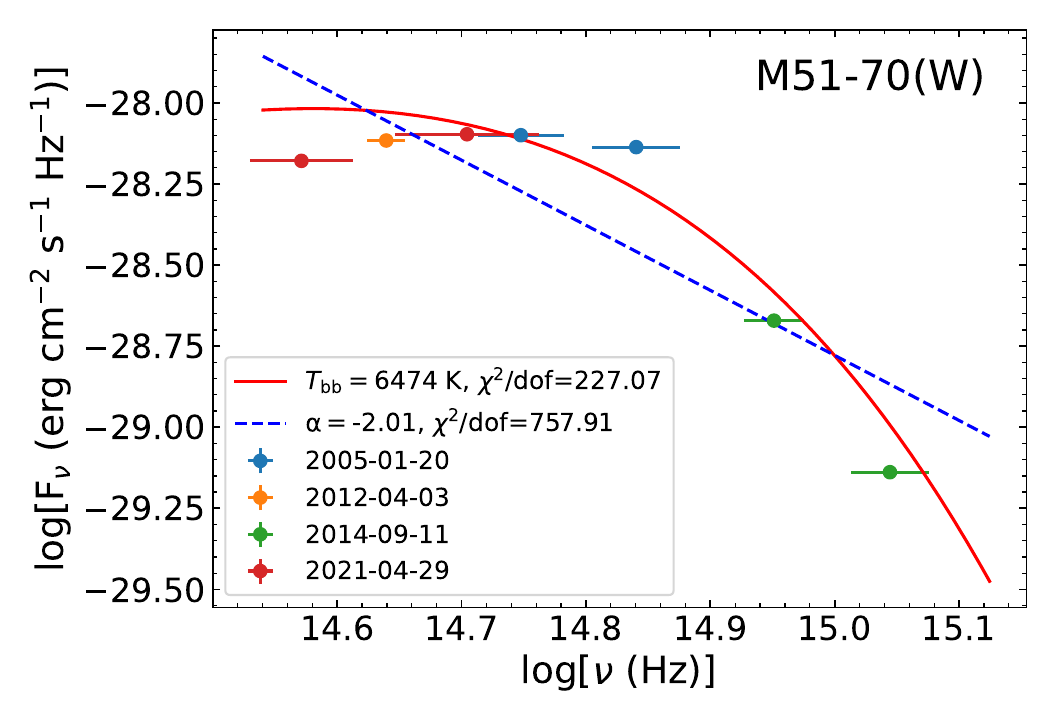}
\includegraphics[width=0.33\linewidth]{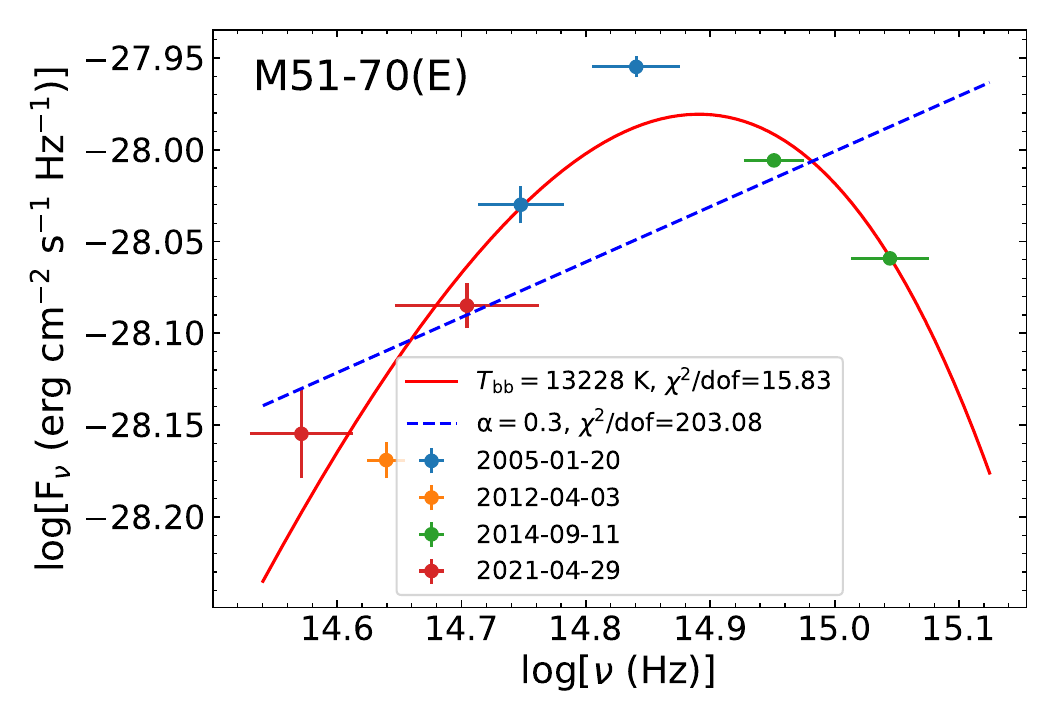}
\includegraphics[width=0.33\linewidth]{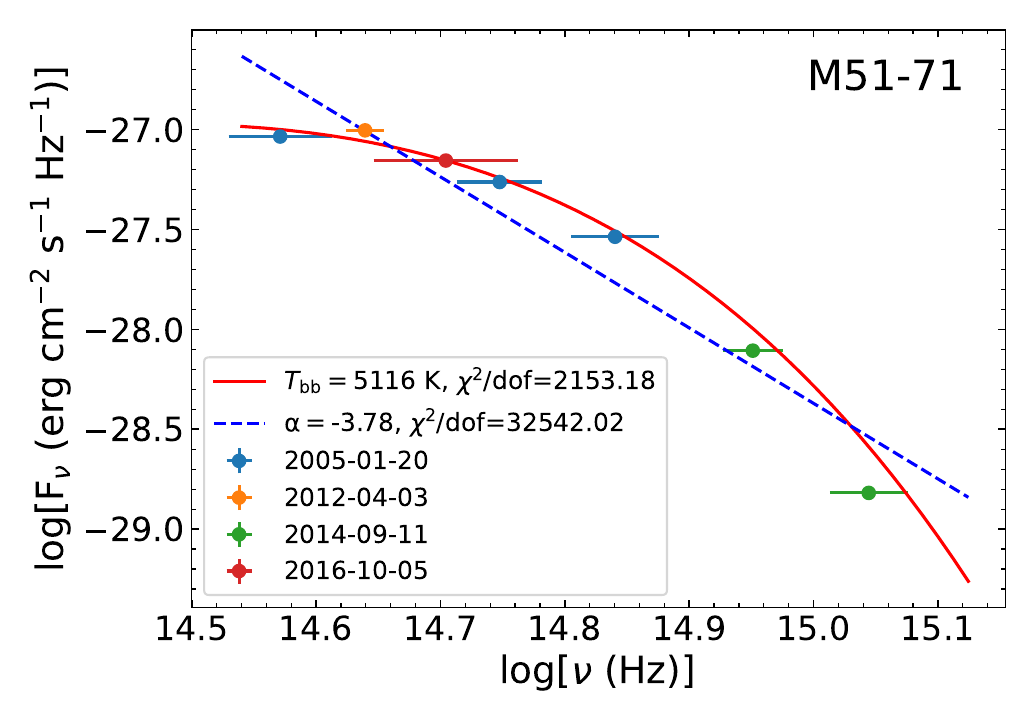}
\includegraphics[width=0.33\linewidth]{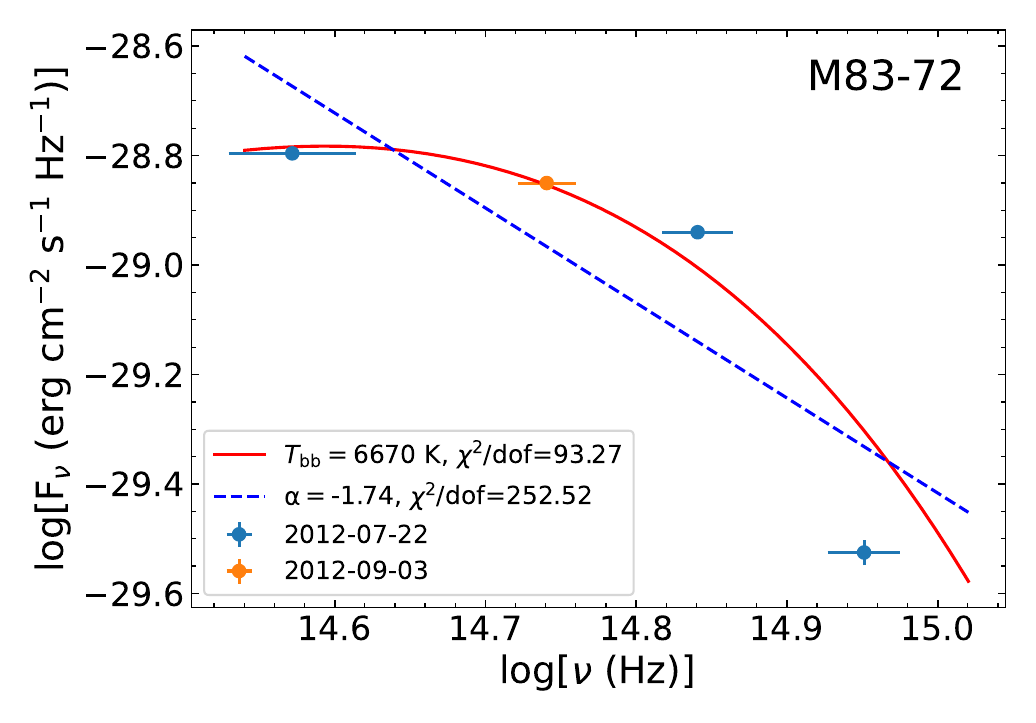}
\includegraphics[width=0.33\linewidth]{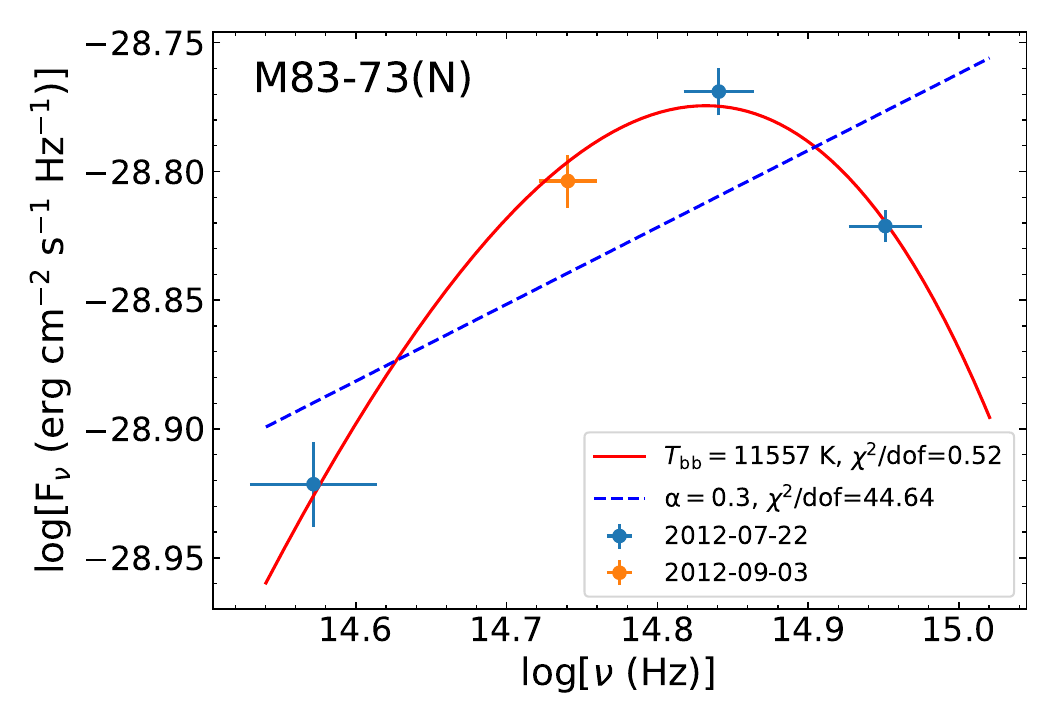}
\includegraphics[width=0.33\linewidth]{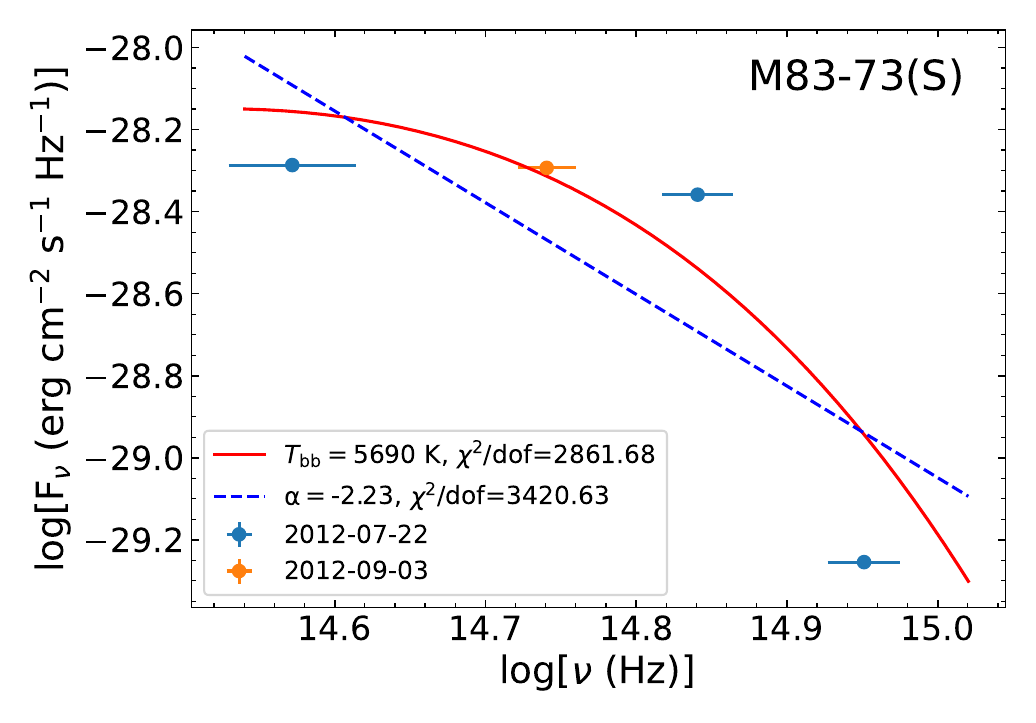}
\includegraphics[width=0.345\linewidth]{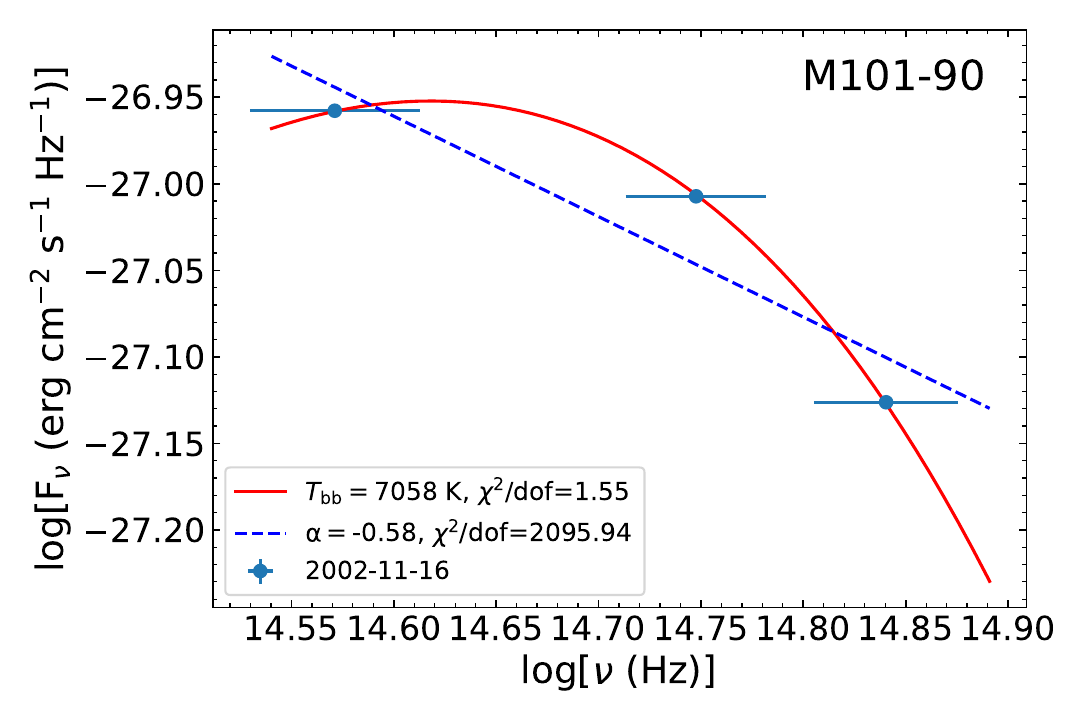}
\caption{SEDs of HST sources with $m < 24$ located out of the corrected 90\% error radius but within the original Chandra 90\% absolute position uncertainty (see Figure~\ref{fig:cp} \& \ref{fig:cp_no}), as a comparison with the identified optical counterparts. The figure is in the same format as that of Figure~\ref{fig:sed}. M51-70 (W) and M51-70 (E) refer to the two sources to the west and east of the original Chandra error circle, respectively. The same for M83-73 (N) and M83-73 (S).}
\label{fig:sed_cat_b}
\end{figure*}
\end{document}